\documentclass[letterpaper]{article} 
\usepackage{aaai2026}  
\usepackage{times}  
\usepackage{amsmath, amssymb}
\usepackage{comment}
\usepackage{multirow}
\usepackage{makecell}
\usepackage{rotating}
\usepackage{helvet}  
\usepackage{courier}  
\usepackage[hyphens]{url}  
\usepackage{graphicx} 
\usepackage{natbib}  
\usepackage{caption} 
\usepackage{algorithm}
\usepackage{algorithmic}
\usepackage{booktabs} 
\usepackage{xcolor}

\usepackage{newfloat}
\usepackage{listings}
\DeclareCaptionStyle{ruled}{labelfont=normalfont,labelsep=colon,strut=off} 
\floatstyle{ruled}
\newfloat{listing}{tb}{lst}{}
\floatname{listing}{Listing}
\newcommand*\samethanks[1][\value{footnote}]{\footnotemark[#1]}

\title{MicroEvoEval: A Systematic Evaluation Framework for Image-Based Microstructure Evolution Prediction}
\author{
  Qinyi Zhang\textsuperscript{\rm 1}\equalcontrib,
  Duanyu Feng\textsuperscript{\rm 2,3}\equalcontrib,
  Ronghui Han\textsuperscript{\rm 1},
  Yangshuai Wang\textsuperscript{\rm 4}\thanks{Co-corresponding author.},
  Hao Wang\textsuperscript{\rm 1}\samethanks
}
\affiliations{
  \textsuperscript{\rm 1} School of Mathematics, Sichuan University, No. 24 Yihuan Road, Chengdu, China.\\
  \textsuperscript{\rm 2} College of Computer Science, Sichuan University, No. 24 Yihuan Road, Chengdu, China.\\
  \textsuperscript{\rm 3} Engineering Research Center of Machine Learning and Industry Intelligence, Ministry of Education, China.\\
  \textsuperscript{\rm 4}Department of Mathematics, National University of Singapore, 10 Lower Kent Ridge Road, Singapore.\\
  
  2024322010008@stu.scu.edu.cn, fengduanyuscu@stu.scu.edu.cn, hronghui@stu.scu.edu.cn, yswang@nus.edu.sg, wangh@scu.edu.cn
}
\usepackage{bibentry}

\begin{document}

\maketitle

\begin{abstract}
Simulating microstructure evolution (MicroEvo) is vital for materials design but demands high numerical accuracy, efficiency, and physical fidelity. Although recent studies on deep learning (DL) offers a promising alternative to traditional solvers, the field lacks standardized benchmarks. Existing studies are flawed due to a lack of comparing specialized MicroEvo DL models with state-of-the-art spatio-temporal architectures, an overemphasis on numerical accuracy over physical fidelity, and a failure to analyze error propagation over time. To address these gaps, we introduce {\tt MicroEvoEval}, the first comprehensive benchmark for image-based microstructure evolution prediction. We evaluate 14 models, encompassing both domain-specific and general-purpose architectures, across four representative MicroEvo tasks with datasets specifically structured for both short- and long-term assessment. Our multi-faceted evaluation framework goes beyond numerical accuracy and computational cost, incorporating a curated set of structure-preserving metrics to assess physical fidelity. Our extensive evaluations yield several key insights. Notably, we find that modern architectures (e.g., VMamba), not only achieve superior long-term stability and physical fidelity but also operate with an order-of-magnitude greater computational efficiency. The results highlight the necessity of holistic evaluation and identify these modern architectures as a highly promising direction for developing efficient and reliable surrogate models in data-driven materials science.
\end{abstract}

\begin{links}
    \link{Code}{https://github.com/ArcueidCroft/MircoEvoEval}
    \link{Datasets}{https://huggingface.co/datasets/ArcueidCroft/MicroEvoEval}
\end{links}

\section{Introduction}
\label{sec:intro}
Material properties are strongly governed by microstructures—the mesoscale arrangement of grains and phases~\cite{olson1997computational, bhadeshia2001importance}. A key objective in materials science is to control microstructural evolution via processing techniques such as casting and annealing~\cite{porterl1992phase, reed2008cambridge}. Accurate prediction of this evolution is critical for designing advanced materials, as it involves complex spatio-temporal dynamics highly sensitive to processing conditions~\cite{seetharaman2021microstructure, chen2002phase, mo2021spatial, jou1997microstructural}. It requires a balance of three key attributes: \textbf{numerical accuracy}, to faithfully capture the evolution of pixel-based states~\cite{sang2023accurate, boettinger2000solidification}; \textbf{computational efficiency}, to enable rapid exploration of design spaces~\cite{tandogan2025multi, noguchi2024microstructure}; and \textbf{physical fidelity}, to ensure that predictions remain consistent with underlying physical laws and preserve essential structural features that are often missed by conventional pixel-based metrics~\cite{kamachali2018numerical, hasan2023microstructure}.

Traditionally, MicroEvo has been modeled using theoretical approaches such as phase-field simulations~\cite{chen2002phase, tourret2022phase, moelans2008introduction, steinbach2013phase}, where the governing partial differential equations (PDEs) are solved using numerical methods like finite difference or spectral methods~\cite{liu2003phase, badalassi2003computation, shen2009efficient}. While accurate, these methods are often computationally expensive and may not be practical for rapid evaluations. Moreover, for complex or poorly understood materials, the PDEs corresponding to the evolution laws may be extremely challenging to formulate. These limitations have motivated the development of data-driven approaches for modeling microstructure evolution.

More recently, deep learning (DL) has opened new avenues for modeling microstructure evolution by learning complex spatio-temporal dynamics directly from image-based data. Initial efforts have shown promise by framing MicroEvo as a image-based sequence prediction task, with evaluations often focused on numerical accuracy (standard image metrics like MSE)~\cite{lanzoni2022morphological, farizhandi2023spatiotemporal}. These studies have frequently adapted established or specialized architectures like E3D-LSTM~\cite{yang2021self} and the recent VMamba~\cite{jing2025research}. However, these models have often been developed in isolation, lacking a systematic comparisons. Concurrently, the broader field of general-purpose spatio-temporal prediction has seen rapid advances, producing powerful models like SimVP.v2~\cite{tan2025simvpv2} and PredFormer~\cite{tang2024predformer}. Despite their success on other diverse tasks, their applicability to the physically-constrained domain of MicroEvo remains largely unevaluated. 

Therefore, a comprehensive benchmark is currently missing to systematically evaluate and compare the diverse deep learning approaches for image-based MicroEvo prediction. This gap is multi-dimensional. Firstly, most evaluations focus on \textbf{numerical accuracy}, while overlooking \textbf{physical fidelity}, the critical aspect for understanding microstructures. Secondly, existing studies rarely offer standardized comparisons between specialized MicroEvo models and potential promising general-purpose spatio-temporal architectures. Despite these, we also find that existing evaluations typically evaluated the long-term accuracy with models trained on short-term data~\cite{yang2021self}. As a result, this methodological gap obscures the process of error accumulation, making it unknown whether a model's short-term accuracy reliably translates to long-term stability.

To address these critical gaps, we present {\tt MicroEvoEval}, the first comprehensive benchmark specifically designed for image-based microstructural evolution. Our approach is built on three key design principles. First, we define a structured suite of tasks and datasets. {\tt MicroEvoEval} includes representative MicroEvo problems such as grain growth and spinodal decomposition, covering a range of physical complexities. The datasets, derived from high-fidelity numerical simulations~\cite{yang2021self}, are further organized into separate test sets that support a joint evaluation of short-term accuracy and long-term stability. This allows for a direct analysis of how errors propagate over time in these systematized tasks. Second, we provide a standardized evaluation across diverse model types. We benchmark 5 models tailored for MicroEvo and 9 general-purpose spatio-temporal architectures, offering the first systematic comparison of their performance under consistent settings. Third, we create a holistic performance assessment. Beyond conventional metrics such as MSE and SSIM for \textbf{numerical accuracy}, our framework incorporates measures of \textbf{physical fidelity} (The Log of Error of Total Area Proportion ``L-ETAP" and The Log of Error of Average Proportion of a Single Region ``L-EAPSR") and \textbf{computational efficiency} (inference time), yielding a more complete and physically meaningful evaluation.

The main contributions of this work are:
(1) \textbf{{\tt MicroEvoEval}: the first standardized benchmark for image-based microstructural evolution.} It comprises representative tasks designed for short- and long-term prediction, along with a comprehensive metric suite that jointly evaluates numerical accuracy, computational efficiency, and physical fidelity.
(2) \textbf{A systematic evaluation across diverse architectural paradigms.} We present the first standardized comparison between MicroEvo-specific models and state-of-the-art general-purpose spatio-temporal architectures, revealing their respective advantages and limitations.
(3) \textbf{Actionable insights for future research.} We demonstrate the necessity of assessing long-term stability and physical fidelity, as short-term numerical accuracy of trained model is a poor indicator. We also show that modern architectures can give a better accuracy-efficiency trade-off, providing a potential path toward developing more reliable and practical models for materials science.

\section{Related Work}
\label{sec:related_work}

\textbf{Physics-Based Numerical Methods.}
Modeling microstructure evolution has traditionally relied on continuum-scale, physics-based formulations. Phase-field models, which describe spatio-temporal dynamics through PDEs for phenomena like solidification and grain growth~\cite{chen2002phase, steinbach1996phase}, are particularly widespread due to their flexibility. However, their high computational cost, especially with explicit time integration schemes, restricts the accessible time and length scales of simulations and hinders rapid exploration of process-parameter space~\cite{greenwood2018quantitative}. Furthermore, for complex or poorly characterized materials, deriving tractable and accurate PDEs can be a significant challenge in itself. While acceleration techniques~\cite{guo2015solving} and alternative data-driven frameworks like Markov Random Fields~\cite{acar2016markov} have been explored to alleviate these issues, the trade-off between fidelity and computational cost remains a primary bottleneck, motivating the search for efficient surrogate models.

\textbf{Deep Learning for Microstructure Evolution.}
To address the computational bottleneck of numerical methods, deep learning (DL) has emerged as a promising alternative for learning MicroEvo dynamics directly from data. Many pioneering studies have adapted established deep learning architectures for this purpose. These approaches often rely on Recurrent Neural Network (RNN) variants to capture temporal dependencies. Examples include adapting classic architectures like ConvLSTM~\cite{mao2024spatiotemporal}, ConvGRU~\cite{lanzoni2022morphological}, and PredRNN~\cite{farizhandi2023spatiotemporal} to forecast microstructural patterns. Other works employ a CNN-RNN structure, using CNNs as powerful feature extractors for the recurrent core. Prominent examples in this category include E3D-LSTM~\cite{yang2021self} and the more recent state-space-based model, VMamba~\cite{jing2025research}. Although these models have demonstrated the feasibility of data-driven MicroEvo prediction, they have been developed and evaluated on disparate tasks and mainly focus on numerical accuracy on long-term prediction, making it difficult to comprehensively compare their relative strengths and weaknesses.

\textbf{General-Purpose Spatio-Temporal Prediction.}
Concurrently, the broader field of computer vision has produced a wealth of powerful models for general-purpose spatio-temporal prediction (i.e., video prediction). These architectures have evolved significantly, from advanced recurrent networks like PredRNN++~\cite{Wang2018predrnn++} and MAU~\cite{chang2021mau} to highly efficient, non-recurrent CNN models such as SimVP~\cite{gao2022simvp} and its successor SimVP.v2~\cite{tan2025simvpv2}. More recently, Transformer-based architectures have become prominent for their ability to capture long-range dependencies, leading to models like PredFormer~\cite{tang2024predformer} and hybrids like SwinLSTM~\cite{tang2023swinlstm}. The cutting edge continues to advance with refined attention mechanisms in models like TAU~\cite{tan2023temporal} and the integration of Mamba in VMRNN~\cite{tang2024vmrnn}. Despite their proven success on different general tasks, these state-of-the-art models are not designed with inherent physical constraints. Their effectiveness in predicting physically plausible microstructure evolution, a domain with fundamentally different underlying rules from natural videos, has not been systematically investigated. Our benchmark aims to bridge this gap by evaluating both domain-specific and general-purpose models under a unified framework.

\section{MicroEvoEval}
\label{sec:microevoeval}

In this section, we present the details of our benchmark {\tt MicroEvoEval}, including the taxonomy of evaluation tasks, the dataset construction, and the metrics used for performance assessment.

\begin{figure*}[ht]
  \centering
  \includegraphics[width=0.9\linewidth]{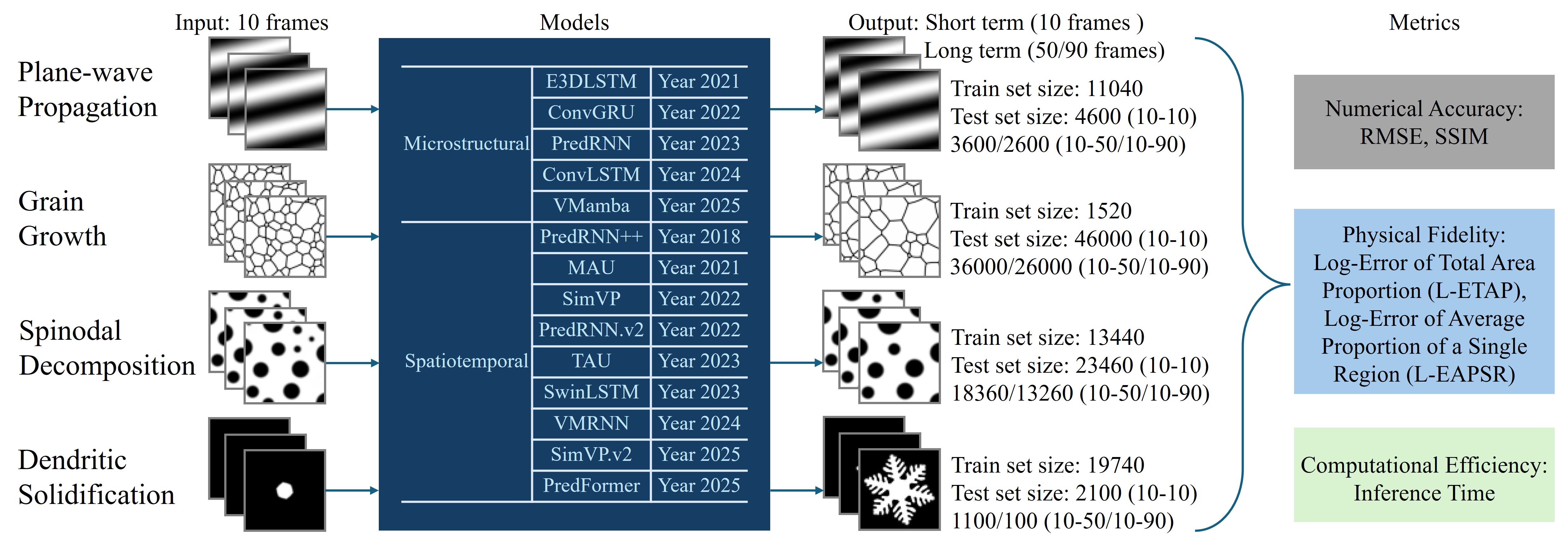}
  \caption{Schematic of the {\tt MicroEvoEval} benchmark for microstructure evolution prediction.}
  \label{Schematic of the Benchmark}
\end{figure*}

\subsection{The Taxonomy of Microstructure Evolution Tasks}



At the microscale, we focus on the evolutions of microstructures governed by known PDEs, as this facilitates robust and quantitative evaluation. These governing PDEs are broadly categorized based on their physical modeling mechanisms.
The first category includes equations describing periodic materials, such as the elastic wave equation for modeling wave propagation. The second involves phase-field models, including the Allen–Cahn (A–C) equation~\cite{allen1979microscopic} for non-conserved order parameters (e.g., crystalline structure, ferromagnetic domains) and the Cahn–Hilliard (C–H) equation~\cite{cahn1958free} for conserved quantities (e.g., composition, mass density). A third category covers multi-physics PDEs that account for coupled interactions among different physical fields. These categories cover all major PDEs governing microstructure evolution, ensuring broad physical representativeness in our study.

In this work, we select representative applications for each of these PDE categories. To ensure systematic coverage, the selected four tasks follow a physical taxonomy encompassing periodic, non-conserved, conserved, and coupled multi-physics mechanisms. Each task was designed to be non-overlapping in its governing dynamics and data domain. \footnote{We adopt PDE-based datasets for their scalability and clean ground truths, as real-world microstructure evolution data remain scarce and noisy for standardized benchmarking.} Specifically, we consider \textbf{plane-wave propagation} for periodic structures, \textbf{grain growth} governed by the A-C equation, \textbf{spinodal decomposition} described by the C–H equation, and \textbf{dendritic solidification} as a coupled multi-physics process. It is worth noting that generating high-quality data for these PDEs is nontrivial. To ensure reproducibility and enable meaningful benchmark comparisons, we adopt the datasets and PDE formulations introduced in~\cite{yang2021self}, which are widely used in the modeling of microstructural evolution~\cite{Fan2024, jing2025research}. We then applied specific processing to these datasets to align them with our evaluation goals in the next subsection. In the following, we first briefly describe each task and present its associated governing equation.\footnote{More details of the physical background and derivation of these tasks are shown in Technical Appendix \ref{app:More Details of the Physical Background}. We also constructed a \textit{Dumbbell} dataset as extended version.}

\begin{itemize}

\item The first task is a simple yet physically relevant toy model that captures the periodic nature of microstructures. It describes the {\bf plane-wave propagation} of a scalar field \( c(x, y, t) \), given by:
\begin{equation}
    c(x, y, t) = \frac{1}{2} \sin(k_x x + k_y y + \omega t + \theta_0) e^{-\beta t} + \frac{1}{2},
    \label{eq:toy_model}
\end{equation}
where \( \vec{k} = (k_x, k_y) \) is the wave vector, \( \theta_0 \) is a random phase, and \( \beta \) is a temporal decay exponent. 

\item The second task focuses on {\bf grain growth}, which is simulated using a multi-order-parameter phase-field model~\cite{moelans2008introduction}, where a set of order parameters \(\{\eta_1(x), \eta_2(x), \ldots, \eta_N(x)\}\) represents \(N\) distinct grain orientations. The total free energy is given by:
\begin{equation}
F = \int \left[ f(\eta_1, \eta_2, \ldots, \eta_N) + \frac{\nu}{2} \sum_{i=1}^{N} (\nabla \eta_i)^2 \right] \,{\rm d}V,
\label{eq:free_energy0}
\end{equation}
where the homogeneous free energy density is defined by~\eqref{eq:grain_local_energy}.
The temporal evolution of each \(\eta_i(x)\) follows the time-dependent A-C equation, a representative {\it second-order} PDE:
\begin{equation}
\frac{\partial \eta_i}{\partial t} = -L \frac{\delta F}{\delta \eta_i}.
\label{eq:allen_cahn}
\end{equation}

\item The third task involves {\bf spinodal decomposition}, a spontaneous phase separation process in binary mixtures. Unlike A–C equation, this process is governed by the {\it fourth-order} C–H equation, which enables domain formation and coarsening without nucleation:
\begin{equation}
\frac{\partial c}{\partial t} = \nabla \cdot \left[ M c (1 - c) \nabla \left( \frac{\partial f_{\text{chem}}}{\partial c} - \epsilon \nabla^2 c \right) \right],
\label{eq:CH}
\end{equation}
where $c$ is the molar fraction of one component in a binary alloy. The homogeneous chemical free energy density is described by~\eqref{eq:regular_solution}.

\item The fourth task models {\bf dendritic solidification}~\cite{kobayashi1993}. The system is described by a temperature field \( T \) and an order parameter \( \varphi \), which distinguishes solid (\( \varphi = 1 \)) and liquid (\( \varphi = 0 \)) phases. The free energy functional is given by:
\begin{equation}
F[\varphi, T] = \int \left[ \frac{1}{2} \epsilon(\theta)^2 |\nabla \varphi|^2 + f(\varphi, T) \right] \mathrm{d} r,
\label{eq:free_energy}
\end{equation}
where anisotropy is introduced via the orientation-dependent gradient coefficient $\epsilon(\theta) = \epsilon_0 \left(1 + \delta \cos[n(\theta - \theta_0)]\right)$, with \( \theta = \arctan(-\varphi_y / \varphi_x) \). The bulk free energy is modeled by a double-well potential from~\eqref{eq:dendrite_bulk}.
The coupled time evolution equations are:
\begin{subequations} \label{eq:dendritic}
\begin{align}
\tau \frac{\partial \varphi}{\partial t} &= -\frac{\delta F}{\delta \varphi}, \label{eq:dendritic_a} \\
\frac{\partial T}{\partial t} &= \nabla^2 T + K \frac{\partial \varphi}{\partial t}, \label{eq:dendritic_b}
\end{align}
\end{subequations}
where \( K \) is the latent heat parameter.

\end{itemize}

\subsection{Tasks and Datasets}
{\tt MicroEvoEval} is built upon these four representative tasks spanning a range of physical phenomena, this section details the design of the tasks and datasets. All underlying data is derived from high-fidelity original simulations on a \(256 \times 256\) grid, subsequently downsampled to a uniform \(64 \times 64\) resolution from original datasets~\cite{yang2021self}.

To investigate the relationship between short-term accuracy and long-term stability, we design a dual evaluation setting based on our curated datasets. The short-term setting assesses a model’s ability to capture immediate dynamics (e.g., predicting the next \texttt{10} frames from \texttt{10} inputs), while the long-term setting evaluates autoregressive performance over extended horizons (up to \texttt{50} and \texttt{90} future frames).\footnote{A key area of focus in this domain is assessing the performance of models when they are trained for short-term prediction but then deployed autoregressively for long-term forecasting. More implementation details are shown in Code and Data Appendix \ref{app:Implementation Details of Datatsets}.} This twofold structure enables a systematic analysis of error accumulation and the extent to which short-term accuracy correlates with long-term reliability. Figure~\ref{Schematic of the Benchmark} summarizes the benchmark tasks, datasets, and key statistics in {\tt MicroEvoEval}.

\paragraph{Plane-wave propagation}
This task utilizes data from simulations of a toy model governed by Equation~\ref{eq:toy_model}, with different $\vec{k}, \omega, \beta, \theta_0$. The original training set providing \texttt{240} sequences of \texttt{200} frames each. For our benchmark, we process their provided training split into \texttt{11040} short clips of \texttt{20} frames to form our training set for forecasting the next \texttt{10} frames given \texttt{10} input frames. We then partition their original test set to create our evaluation sets. The short-term prediction task has \texttt{4600} test cases. For the long-term prediction task predicting the next 50 frames, we make \texttt{3600} test cases, and \texttt{2600} test cases for the task predicting next 90 frames.

\paragraph{Grain growth}
The dataset for this task is based on simulations of polycrystalline grain growth governed by the Allen-Cahn Equation~\eqref{eq:allen_cahn}. The initial polycrystalline structures were generated via Voronoi tessellation with 100 random seeds. The source data consists of \texttt{200}-frame sequences depicting grain coarsening. We adopt the designated training clips from the original study to form our training set of \texttt{1520} samples. For our evaluation sets, the short-term task involves \texttt{46000} test cases, while the long-term task involves \texttt{36000} test cases for the task predicting next 50 frames and \texttt{26000} test cases for the task predicting next 90 frames.

\paragraph{Spinodal decomposition}
This task is based on simulations solving the Cahn-Hilliard Equation~\eqref{eq:CH}  with varying initial configurations sampled from \( c_0 = 0.25, 0.5, \) or \( 0.75 \), plus random noise \( \Delta c = 0.01 \) to initiate phase separation $c$ . The original work generated \texttt{640} simulations, each \texttt{100} frames long. Following their methodology, we use staggered clips from their training simulations to create our \texttt{13440} training samples. The test simulations are then used by us to construct distinct short-term (\texttt{23460} test cases) and long-term (\texttt{18360} test cases) evaluation sets for task predicting next 50 frames while the long-term task (predicting next 90 frames) involves \texttt{13260} test cases.

\paragraph{Dendritic Solidification}
For this task, we employ datasets from simulations of dendritic solidification based on Equation~\eqref{eq:dendritic}. The original study provides \texttt{940} training simulations, each producing \texttt{100} frames. Our training set is built from overlapping clips extracted from these simulations, resulting in \texttt{19740} samples. The provided test simulations, which notably include out-of-distribution samples to assess generalization, are partitioned by us into evaluation sets for short-term (\texttt{2100} test cases) and long-term (\texttt{1100} test cases for predicting next 50 frames and 100 test cases for predicting next 90 frames) performance assessment.

\begin{table*}[htbp]
\renewcommand{\arraystretch}{0.55}
\setlength{\tabcolsep}{0.5mm} 
  \centering
  \small
  \caption{The performance of different models in short-term prediction (10-10) of {\tt MicroEvoEval}.}
    \begin{tabular}{cccccccccccccc}
    \toprule
          &       & \multicolumn{3}{c}{Plane-wave propagation} & \multicolumn{3}{c}{Grain growth} & \multicolumn{3}{c}{Spinodal decomposition} & \multicolumn{3}{c}{Dendritic solidification} \\
    \midrule
    Model domain & Model & RMSE  & SSIM  & L-ETAP & RMSE  & SSIM  & L-EASPR & RMSE  & SSIM  & L-ETAP & RMSE  & SSIM  & L-ETAP \\
    \midrule
    \multicolumn{1}{c}{\multirow{5}[1]{*}{Microstructural}} & E3DLSTM & 0.01673 & 0.99315 & -0.993 & 0.03394 & 0.98607 & -2.488 & 0.00418 & 0.99970 & -2.826 & 0.01275 & 0.99717 & -2.540 \\
          & ConvGRU & 0.00338 & 0.99944 & -0.994 & 0.02134 & 0.99183 & -2.586 & 0.00389 & 0.99967 & -2.853 & 0.00267 & 0.99965 & -2.865 \\
          & PredRNN & 0.00102 & 0.99994 & -0.995 & 0.02453 & 0.99165 & -2.541 & \textbf{0.00270} & 0.99985 & \textbf{-2.916} & 0.00574 & 0.99911 & -2.737 \\
          & ConvLSTM & 0.00328 & 0.99965 & \textbf{-0.996} & 0.03082 & 0.98707 & -2.492 & 0.00423 & 0.99969 & -2.857 & 0.00545 & 0.99912 & -2.726 \\
          & VMamba & 0.00151 & 0.99991 & -0.993 & \textbf{0.00871} & \textbf{0.99926} & \textbf{-2.875} & 0.00382 & 0.99983 & -2.902 & \textbf{0.00259} & \textbf{0.99990} & \textbf{-2.942} \\
    \midrule
    \multicolumn{1}{c}{\multirow{9}[1]{*}{Spatiotemporal}} & PredRNN++ & 0.00108 & 0.99993 & -0.993 & 0.03071 & 0.98795 & -2.545 & 0.00285 & 0.99981 & -2.844 & 0.00662 & 0.99883 & -2.702 \\
          & MAU   & 0.00867 & 0.99732 & -0.977 & 0.05939 & 0.95131 & -1.830 & 0.00773 & 0.99876 & -2.494 & 0.00752 & 0.99828 & -2.553 \\
          & SimVP & 0.00471 & 0.99905 & -0.993 & 0.02563 & 0.99139 & -2.600 & 0.00492 & 0.99961 & -2.678 & 0.00646 & 0.99887 & -2.665 \\
          & PredRNN.v2 & 0.00157 & 0.99986 & -0.993 & 0.03954 & 0.97898 & -2.421 & 0.00425 & 0.99964 & -2.844 & 0.00561 & 0.99915 & -2.739 \\
          & TAU   & 0.00271 & 0.99942 & -0.992 & 0.02127 & 0.99463 & -2.633 & 0.00421 & 0.99973 & -2.864 & 0.00441 & 0.99954 & -2.741 \\
          & SwinLSTM & 0.00236 & 0.99938 & -0.995 & 0.03269 & 0.98502 & -2.570 & 0.00354 & 0.99932 & -2.868 & 0.00764 & 0.99877 & -2.612 \\
          & VRMNN & \textbf{0.00081} & \textbf{0.99998} & -0.995 & 0.02230 & 0.99602 & -2.618 & 0.00347 & \textbf{0.99987} & -2.843 & 0.00668 & 0.99962 & -2.673 \\
          & SimVP.v2 & 0.00578 & 0.99920 & -0.993 & 0.02366 & 0.99339 & -2.603 & 0.00442 & 0.99954 & -2.755 & 0.00495 & 0.99945 & -2.638 \\
          & PredFormer & 0.00917 & 0.98603 & -0.972 & 0.02453 & 0.99165 & -2.696 & 0.01273 & 0.96767 & -2.243 & 0.01261 & 0.95548 & -2.443 \\
    \bottomrule
    \end{tabular}%
  \label{tab:short_term_results}%
\end{table*}%

\subsection{Metrics}

To enable a holistic assessment, our evaluation framework integrates metrics across three dimensions: \textbf{predictive accuracy}, \textbf{physical fidelity}, and \textbf{computational efficiency}, each critical for practical applications in materials science.\footnote{More implementation details are shown in Code and Data Appendix \ref{app:Implementation Details of metrics}.}

\paragraph{Numerical Accuracy}
It is assessed using standard image similarity metrics that quantify pixel-wise differences between the predicted and ground truth frames. We employ two widely-used metrics used in previous studies~\cite{yang2021self,farizhandi2023spatiotemporal}: the Root Mean Squared Error (RMSE) and the Structural Similarity Index Measure (SSIM). For each frame, RMSE is defined as:
\begin{equation}
\text{RMSE} = \sqrt{ \frac{1}{N_x N_y} \sum_{i=1}^{N_x} \sum_{j=1}^{N_y} \left(p_g(i,j) - p_p(i,j)\right)^2 },
\label{eq:rmse}
\end{equation}
where $p_g(i,j)$ and $p_p(i,j)$ are the pixel values of the ground truth $g$ and prediction $p$, respectively. $N_x, N_y$ are the width and height of the image in pixels. For each frame, SSIM provides a measure of perceived structural similarity and is defined as:
\begin{equation}
\label{eq:ssim}
\text{SSIM} = 
\frac{
\left(2\bar{p}_g \bar{p}_p + c_1\right)\left(2\sigma_{gp} + c_2\right)
}{
\left(\bar{p}_g^2 + \bar{p}_p^2 + c_1\right)\left(\sigma_g^2 + \sigma_p^2 + c_2\right)
},\
\end{equation}
where $\bar{p}_k$ and $\sigma_k$ ($k = g, p$) are the average pixel value and variance, and $\sigma_{gp}$ is their covariance. The constants $c_1$ and $c_2$ stabilize the division. The final metric scores are computed by averaging the frame-wise results over all predicted frames and test samples.

\paragraph{Physical Fidelity}
Recognizing that pixel-level metrics may not adequately capture the preservation of essential physical properties, we introduce two custom metrics to assess physical fidelity. We define that $\Omega_{i,t}$ is the frame at time $t$ for sample $i =1,\cdots, N$.


For tasks where the overall phase fraction is a key evolving property, such as plane-wave propagation, spinodal decomposition, and dendritic solidification, we introduce the \textbf{Log-Error of Total Area Proportion (L-ETAP)}. This metric is physically significant as it evaluates a model's ability to conserve the total mass or volume of the evolving phase, a fundamental physical constraint. Specifically, we track this by computing the phase area fraction $A(\Omega_{i,t})/S$, where $A(\Omega_{i,t})$ is the total area of the evolving phase (e.g., the white pixels for dendrites) and $S$ is the total image area. Any deviation in this fraction indicates unphysical mass gain or loss over time. L-ETAP therefore captures the degree to which a model maintains global conservation laws across a temporal sequence. It is defined as:
\begin{equation}
\label{eq:l-etap}
    \text{L-ETAP} = \log_{10} \left( \sqrt{\frac{1}{N}\sum_{i=1}^{N}  \sum_{t=0}^{T}\left(  \frac{A(\Omega^g_{i,t})}{S}-\frac{A(\Omega^p_{i,t})}{S}\right)^2} \right).
\end{equation}


For the grain growth task, where the kinetics of grain coarsening are critical, we introduce the \textbf{Log-Error of Average Proportion of a Single Region (L-EAPSR)}. This metric's physical meaning lies in its ability to assess whether a model accurately captures the change in average grain size by measuring the error in the number of grains over time. $C(\Omega_{i,t})$ counts the number of grains (connected regions) at each frame, the reciprocal $1/C$ approximates the average area of a single grain. 
L-EAPSR thus reflects how well the model captures this key dynamical feature of the underlying phase-field evolution.
It is defined as:
\begin{equation}
\label{eq:l-eapsr}
    \text{L-EAPSR} = \log_{10} \left( \sqrt{\frac{1}{N}\sum_{i=1}^{N}  \sum_{t=0}^{T}\left(  \frac{1}{C(\Omega^g_{i,t})}-\frac{1}{C(\Omega^p_{i,t})}\right)^2} \right).
\end{equation}
Consequently, the final values for both metrics will be less than zero. A result that is more negative (more smaller), signifies that the model better maintains the physical properties.

\begin{table*}[htbp]
\renewcommand{\arraystretch}{0.55}
\setlength{\tabcolsep}{0.5mm} 
  \centering
  \small
  \caption{The performance of different models in long-term prediction (10-50/10-90) of {\tt MicroEvoEval}.}
    \begin{tabular}{cccccccccccccc}
\toprule        &       & \multicolumn{3}{c}{Plane-wave propagation} & \multicolumn{3}{c}{Grain growth} & \multicolumn{3}{c}{Spinodal decomposition} & \multicolumn{3}{c}{Dendritic solidification} \\
\midrule
    Model domain & Model & RMSE  & SSIM  & L-ETAP & RMSE  & SSIM  & L-EASPR & RMSE  & SSIM  & L-ETAP & RMSE  & SSIM  & L-ETAP \\
    \midrule
    \multicolumn{14}{c}{Long term prediction (10-50)} \\
    \midrule
    \multicolumn{1}{c}{\multirow{5}[1]{*}{Microstructural}} & E3DLSTM & 0.11997 & 0.78022 & -0.667 & 0.11065 & 0.87022 & -1.750 & 0.01951 & 0.99546 & -1.809 & 0.05015 & 0.97352 & -1.188 \\
          & ConvGRU & 0.01845 & 0.98973 & -0.714 & 0.08843 & 0.91765 & -1.760 & 0.01691 & 0.99727 & -2.139 & 0.02720 & 0.99092 & -1.708 \\
          & PredRNN & 0.00819 & 0.99742 & \textbf{-0.730} & 0.22172 & 0.46442 & -0.685 & 0.01758 & 0.99655 & -1.817 & 0.25915 & 0.71776 & -0.032 \\
          & ConvLSTM & 0.11670 & 0.73850 & -0.622 & 0.10027 & 0.89229 & -1.564 & 0.02520 & 0.99230 & -1.783 & 0.04143 & 0.98008 & -1.278 \\
          & VMamba & 0.00817 & 0.99709 & -0.728 & \textbf{0.02986} & \textbf{0.99008} & \textbf{-2.283} & \textbf{0.01342} & \textbf{0.99848} & \textbf{-2.221} & \textbf{0.01922} & \textbf{0.99609} & \textbf{-1.983} \\
           \midrule
    \multicolumn{1}{c}{\multirow{9}[1]{*}{Spatiotemporal}} & PredRNN++ & \textbf{0.00754} & \textbf{0.99803} & -0.704 & 0.09926 & 0.89755 & -1.764 & 0.02294 & 0.99393 & -1.916 & 0.04285 & 0.97866 & -1.320 \\
          & MAU   & 0.15833 & 0.65653 & -0.411 & 0.16699 & 0.68662 & -1.093 & 0.05056 & 0.96809 & -1.144 & 0.09578 & 0.92582 & -0.813 \\
          & SimVP & 0.02054 & 0.98538 & -0.725 & 0.09435 & 0.89925 & -1.899 & 0.02279 & 0.99565 & -1.820 & 0.03226 & 0.98725 & -1.777 \\
          & PredRNN.v2 & 0.01160 & 0.99545 & -0.708 & 0.16927 & 0.48566 & -0.487 & 0.02493 & 0.99300 & -1.510 & 0.04015 & 0.98058 & -1.402 \\
          & TAU   & 0.01396 & 0.99389 & -0.727 & 0.08250 & 0.92942 & -1.868 & 0.01643 & 0.99809 & -2.213 & 0.02374 & 0.99233 & -1.797 \\
          & SwinLSTM & 0.02046 & 0.98421 & -0.726 & 0.12624 & 0.81917 & -1.805 & 0.01745 & 0.99635 & -2.000 & 0.04339 & 0.98029 & -1.513 \\
          & VRMNN & 0.01113 & 0.99683 & -0.728 & 0.06248 & 0.96615 & -1.975 & 0.01888 & 0.99753 & -1.911 & 0.03521 & 0.98984 & -1.651 \\
          & SimVP.v2 & 0.02234 & 0.98652 & -0.724 & 0.08931 & 0.91645 & -1.851 & 0.01950 & 0.99679 & -2.063 & 0.02615 & 0.99109 & -1.762 \\
          & PredFormer & 0.02892 & 0.95577 & -0.695 & 0.08450 & 0.91212 & -2.022 & 0.05575 & 0.93378 & -1.332 & 0.05868 & 0.92531 & -1.380 \\
    \midrule
    \multicolumn{14}{c}{Long term prediction (10-90)} \\
    \midrule
    \multicolumn{1}{c}{\multirow{5}[1]{*}{Microstructural}} & E3DLSTM & 0.17089 & 0.58738 & -0.489 & 0.14800 & 0.76496 & -1.450 & 0.03642 & 0.98552 & -1.441 & 0.16485 & 0.84943 & -0.212 \\
          & ConvGRU & 0.03635 & 0.95884 & -0.596 & 0.12688 & 0.82782 & -1.425 & 0.02791 & 0.99304 & -1.843 & 0.08216 & 0.95175 & -1.070 \\
          & PredRNN & 0.02630 & 0.97109 & -0.598 & 0.31882 & 0.26338 & 0.060 & 0.03744 & 0.98465 & -1.301 & 0.42713 & 0.43134 & 0.296 \\
          & ConvLSTM & 0.16098 & 0.54013 & -0.442 & 0.13749 & 0.79176 & -1.173 & 0.05765 & 0.96745 & -1.268 & 0.12132 & 0.90480 & -0.489 \\
          & VMamba & \textbf{0.01652} & \textbf{0.98745} & -0.629 & \textbf{0.04727} & \textbf{0.97322} & \textbf{-2.011} & \textbf{0.02138} & \textbf{0.99627} & -1.944 & \textbf{0.05542} & \textbf{0.97682} & \textbf{-1.437} \\
           \midrule
    \multicolumn{1}{c}{\multirow{9}[0]{*}{Spatiotemporal}} & PredRNN++ & 0.02281 & 0.98039 & \textbf{-0.633} & 0.13494 & 0.80531 & -1.458 & 0.05207 & 0.97385 & -1.437 & 0.12542 & 0.90127 & -0.485 \\
          & MAU   & 0.18475 & 0.51672 & -0.340 & 0.20358 & 0.53944 & -0.871 & 0.09445 & 0.92043 & -0.760 & 0.18592 & 0.82421 & -0.349 \\
          & SimVP & 0.03558 & 0.95929 & -0.621 & 0.13397 & 0.80088 & -1.605 & 0.03582 & 0.98976 & -1.556 & 0.07325 & 0.95754 & -1.179 \\
          & PredRNN.v2 & 0.03477 & 0.96081 & \textbf{-0.633} & 0.20170 & 0.28880 & -0.492 & 0.05702 & 0.96639 & -0.888 & 0.11749 & 0.91178 & -0.587 \\
          & TAU   & 0.02623 & 0.97736 & -0.626 & 0.12098 & 0.84532 & -1.557 & 0.02517 & 0.99590 & \textbf{-1.958} & 0.05931 & 0.96924 & -1.209 \\
          & SwinLSTM & 0.05263 & 0.91687 & -0.594 & 0.17355 & 0.66420 & -1.408 & 0.03240 & 0.98954 & -1.620 & 0.09784 & 0.93640 & -0.956 \\
          & VRMNN & 0.03528 & 0.96877 & -0.615 & 0.09091 & 0.91929 & -1.697 & 0.03202 & 0.99288 & -1.524 & 0.09203 & 0.94758 & -1.140 \\
          & SimVP.v2 & 0.03878 & 0.95755 & -0.617 & 0.12863 & 0.82407 & -1.536 & 0.03004 & 0.99268 & -1.811 & 0.06307 & 0.96618 & -1.149 \\
          & PredFormer & 0.05832 & 0.87576 & -0.560 & 0.12084 & 0.82722 & -1.741 & 0.09589 & 0.89108 & -0.990 & 0.13028 & 0.85560 & -0.823 \\
          \bottomrule
    \end{tabular}%
  \label{tab:long_term_results}%
\end{table*}%

\paragraph{Computational Efficiency}
Finally, to evaluate the practical applicability of each model, we measure its computational efficiency. This is quantified by the average inference time required to predict a complete short-term sequence (forecasting \texttt{10} frames from \texttt{10} input frames).\footnote{Since the computational time for autoregressive forecasting scales linearly with the number of predicted frames, we report this short-term inference time as a representative measure of a model's efficiency.}

\section{Experiment}
\label{sec:expriment}


\subsection{Experimental Setup}

To conduct this evaluation, we assess a diverse set of 14 models, comprising 5 architectures specifically tailored for MicroEvo and 9 state-of-the-art general-purpose spatio-temporal models. The domain-specific MicroEvo models include E3D-LSTM~\cite{yang2021self},
ConvGRU~\cite{lanzoni2022morphological},
PredRNN~\cite{farizhandi2023spatiotemporal},
ConvLSTM~\cite{mao2024spatiotemporal}, and
VMamba~\cite{jing2025research}. The general-purpose spatio-temporal architectures consist of PredRNN++~\cite{Wang2018predrnn++},
MAU~\cite{chang2021mau},
SimVP~\cite{gao2022simvp},
PredRNN.v2~\cite{wang2022predrnn},
TAU~\cite{tan2023temporal},
SwinLSTM~\cite{tang2023swinlstm},
VMRNN~\cite{tang2024vmrnn},
SimVP.v2~\cite{tan2025simvpv2}, and
PredFormer~\cite{tang2024predformer}. 
All models were trained for a maximum of 200, 300, 300, and 400 epochs of each task with the best performing checkpoint selected based on performance in a held-out validation set.\footnote{Further implementation details of each model can be found in Code and Data Appendix~\ref{app:implementation_details_of_model}.} 

\subsection{Main Results}
Our experiments provide a deep understanding of MicroEvo prediction by investigating four core questions: (1) whether a model's \textbf{short-term accuracy} is a reliable indicator of its \textbf{long-term stability}; (2) whether standard image metrics (numerical accuracy) are sufficient for evaluation compared to our curated metrics for \textbf{physical fidelity}; (3) how domain-specific MicroEvo models perform relative to general-purpose spatiotemporal architectures; and (4) which architectural classes offer the best trade-off between performance and \textbf{computational efficiency}.\footnote{Additional accuracy metrics, performance-cost figure, statistical variance analyses, cases and out-of-distribution evaluations are provided in Technical Appendix~\ref{app:More Results}.}

\textbf{Short-term performance is a poor indicator of long-term stability.}
Comparing Table~\ref{tab:short_term_results} and Table~\ref{tab:long_term_results}, nearly all models exhibit a significant degradation in performance during long-term autoregressive forecasting. For instance, while VMRNN achieves the best short-term performance on the Plane-wave Propagation task, its advantage diminishes significantly in long-term prediction. Another example is PredRNN, which, despite its strong short-term result on Spinodal Decomposition, suffers a catastrophic performance collapse on the long-term Grain Growth task (SSIM of 0.263), demonstrating severe error accumulation. In contrast, VMamba consistently maintains top-tier performance across both short- and long-term horizons, especially in the most challenging 90-frame predictions, highlighting its superior stability. Therefore, for MicroEvo, the results from short-term training are unreliable, the long-term evaluation is essential for identifying robust models.

\textbf{Physical fidelity metrics provide indispensable insights beyond standard image metrics.}
This is evidenced by frequent divergences where the model that performs best for numerical accuracy such as RMSE and SSIM is not the best performer on our physical fidelity metrics. For instance, in the 90-frame long-term prediction for Spinodal Decomposition, VMamba achieves the best numerical accuracy with the lowest RMSE (0.02138) and highest SSIM (0.99627). However, the best physical fidelity, as measured by L-ETAP, is achieved by a different model, TAU (-1.958). This divergence illustrates that minimizing pixel-level error does not ensure optimal adherence to the underlying physical constraints of the system, such as phase fraction conservation. Therefore, relying solely on standard vision metrics is insufficient, and physical fidelity metrics are necessary to ensure that predictions are physically meaningful.

\textbf{The architecture of model is more critical than domain-specific and general-purpose methods.} The overall top-performing model is VMamba, a Microstructural model that leverages a modern architecture (Mamba). It consistently excels, particularly in long-term predictions and complex tasks. However, another top model, especially in short-term tasks, is VMRNN, a Spatiotemporal model that also integrates a Mamba-like structure. The common thread between these top performers is their advanced architecture, which contrasts with the generally weaker long-term stability of older RNN-based models from both categories (e.g., PredRNN, ConvLSTM). This suggests the most promising path forward is combining state-of-the-art architectural designs with domain-specific considerations.

\begin{table}[htbp]
\renewcommand{\arraystretch}{0.75}
\setlength{\tabcolsep}{0.8mm} 
  \centering
  \small
  \caption{The time of different models in short-term prediction (10-10) of {\tt MicroEvoEval}.}
    \begin{tabular}{ccccc}
    \toprule
    Model & 
    \thead{Plane-wave \\ propagation} & 
    \thead{Grain \\ growth} & 
    \thead{Spinodal \\ decomposition} & 
    \thead{Dendritic \\ solidification}  \\
    \midrule
    E3DLSTM & 0.113 & 0.103 & 0.111 & 0.113 \\
    ConvGRU & 0.044 & 0.195 & 0.039 & 0.044 \\
    PredRNN & 0.096 & 0.087 & 0.097 & 0.096 \\
    ConvLSTM & 0.095 & 0.089 & 0.097 & 0.095 \\
    VMamba & \textbf{0.021} & \textbf{0.006} & \textbf{0.007} & \textbf{0.021} \\
    \midrule
    PredRNN++ & 0.093 & 0.088 & 0.097 & 0.098 \\
    MAU   & 0.087 & 0.093 & 0.090 & 0.094 \\
    SimVP & 0.090 & 0.084 & 0.096 & 0.094 \\
    PredRNN.v2 & 0.088 & 0.088 & 0.095 & 0.097 \\
    TAU   & 0.086 & 0.082 & 0.090 & 0.097 \\
    SwinLSTM & 0.097 & 0.098 & 0.105 & 0.106 \\
    VMRNN &     0.044  &  0.034     & 0.035      &0.039  \\
    SimVP.v2 & 0.102 & 0.083 & 0.091 & 0.093 \\
    PredFormer & 0.085 & 0.083 & 0.086 & 0.089 \\
    \bottomrule
    \end{tabular}%
  \label{tab:inference_time}%
\end{table}%

\textbf{VMamba not only delivers state-of-the-art performance but also operates with an order-of-magnitude greater efficiency than most other models.} Shown in Table \ref{tab:inference_time}, with an inference time of just 0.006s for the Grain Growth task, compared to the 0.08-0.1s range for most competitors, VMamba shows great potential of a trade-off between accuracy and speed. Its superior performance and efficiency make it a highly promising candidate for practical applications. This further highlights the transformative potential of modern architectures like Mamba.

\subsection{Visual Analysis}

\begin{figure}[ht]
  \centering
  \includegraphics[width=0.9\linewidth]{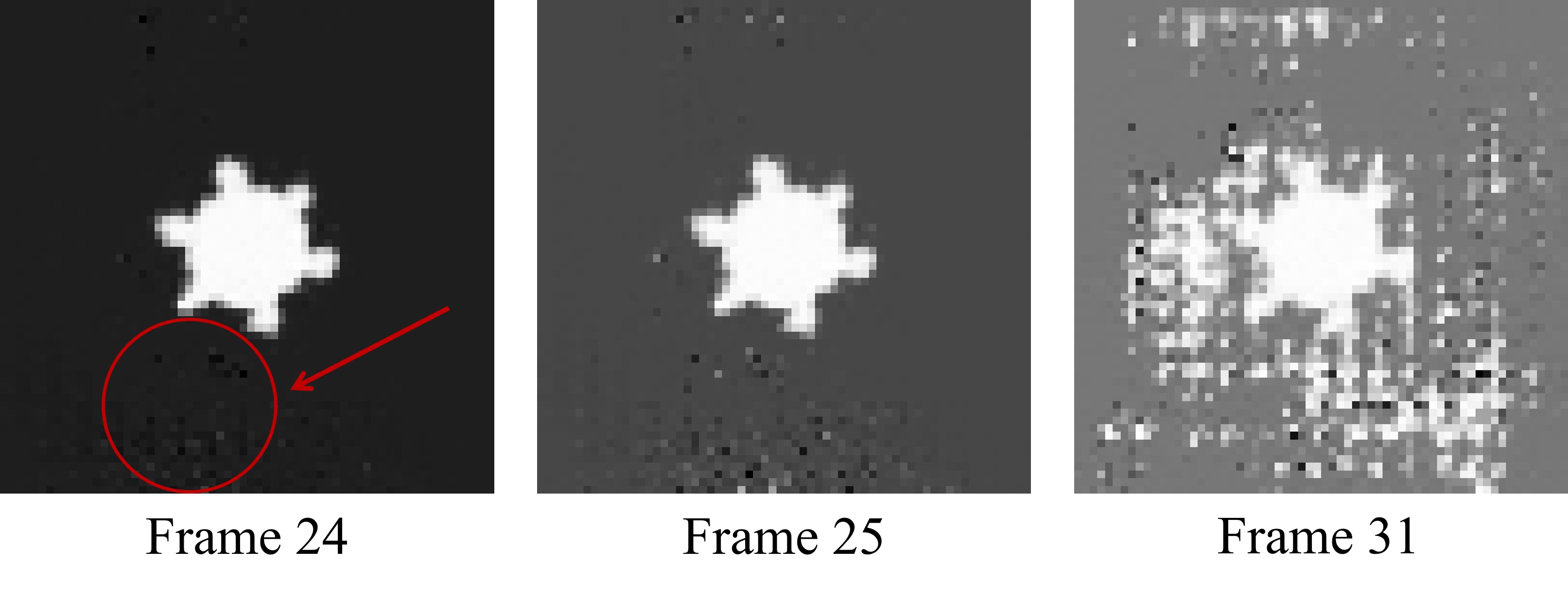}
  \caption{Case study on dendritic solidification.}
  \label{Case Study on Dendritic Solidification.}
\end{figure}

In addition to quantitative metrics, we use visual analysis to explore long-term stability and to show the findings of our physical fidelity metrics.

Figure \ref{Case Study on Dendritic Solidification.} shows why short-term accuracy is a poor indicator of long-term stability. Using the Dendritic Solidification task as an example, we observe that in an early prediction frame (Frame 24), a model like PredRNN introduces low-amplitude noise artifacts into the background, as highlighted by the red circle. Although this initial error may be minor, its effect becomes catastrophic during forecasting. This leads to a complete breakdown of the physical structure in Frame 31, where the prediction has diverged into a meaningless state.

\begin{figure}[ht]
  \centering
  \includegraphics[width=0.92\linewidth]{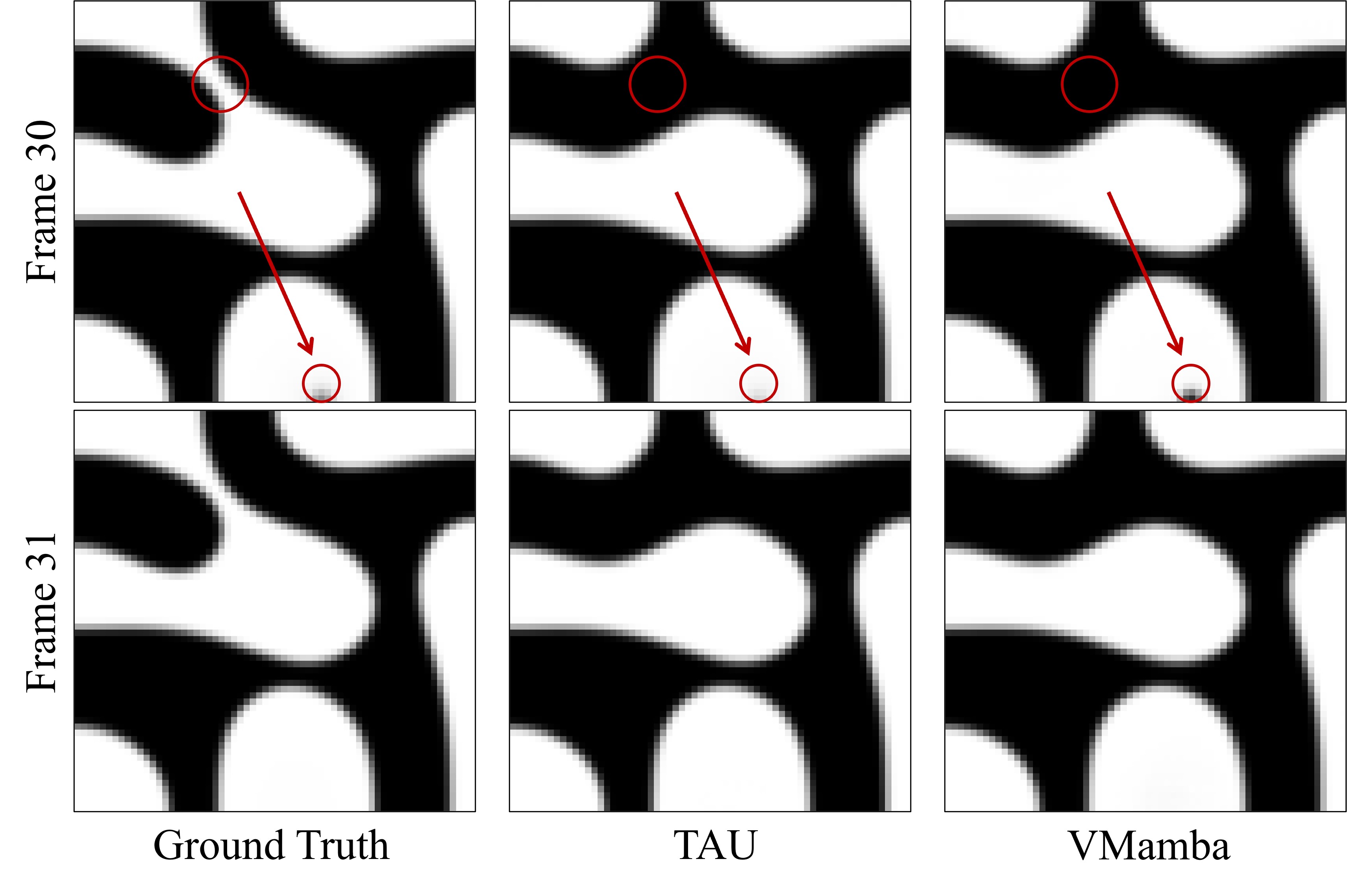}
  \caption{Case study on spinodal decomposition. }
  \label{Case Study on Spinodal Decomposition.}
\end{figure}

Figure \ref{Case Study on Spinodal Decomposition.} shows how our physical fidelity metrics can further distinguish between two strong models. Both models achieve high numerical scores, including the eventual dissolution of the smaller droplet highlighted by the red arrow. However, they differ in the speed of this process. The ground truth shows the droplet gradually shrinking from Frame 30 to Frame 31. TAU's prediction closely mimics this gradual dissolution rate. In contrast, VMamba predicts an overly accelerated dynamic where the droplet disappears more rapidly than in the ground truth.

\section{Conclusion}
In this work, we introduced {\tt MicroEvoEval}, the first comprehensive benchmark to address critical gaps in evaluating deep learning for microstructure evolution. Our framework systematically compares domain-specific and general-purpose models using a multi-faceted evaluation of short- and long-term stability, numerical accuracy, physical fidelity and efficiency. Our results demonstrate the necessity of both long-term evaluation and physical fidelity metrics, as short-term numerical accuracy is a poor indicator of overall performance. Furthermore, we show that modern architectures, particularly state-space models, deliver superior performance and stability. These findings provide guidance for future research, pointing towards the development of more robust, physics-informed, and computationally efficient models for materials science.
\footnote{We discuss the other future directions and limitations of our work in Technical  Appendix \ref{app:limitation}.}

\section*{Acknowledgements}
This research is supported by the National Natural Science Foundation of China (No. 11971336, 12571435).

\bibliography{aaai2026}

\clearpage
\newpage
\appendix

\newcommand{\zqy}[1]{{\color{magenta} #1}}
\newcommand{\czqy}[1]{{\color{magenta} [QY: #1]}}

This is the \textbf{Technical Appendix} of our paper. In this part, we will give more details of the physical background of the tasks in our benchmark, more detailed results of experience, and the limitations of our work.

\section{More Details of the Physical Background}
\label{app:More Details of the Physical Background}

\paragraph{Time-Decaying Plane Wave Propagation} As an easy yet physically relevant test case, we consider a benchmark involving the propagation of a two-dimensional time-decaying plane wave. The scalar field $c(x, y, t)$ is defined analytically as:
\begin{equation}
c(x, y, t) = \frac{1}{2} \sin(k_x x + k_y y + \omega t + \theta_0) e^{-\beta t} + \frac{1}{2},
\label{eq:plane_wave_decay}
\end{equation}
where $(k_x, k_y)$ is the wave vector that determines the spatial direction and wavelength of the plane wave, $\omega$ is the temporal frequency, $\theta_0 \sim \mathcal{U}(0, 2\pi)$ is a random initial phase, and $\beta > 0$ is the temporal decay coefficient. The additive constant $+1/2$ ensures non-negativity of the scalar field, which is desirable when modeling physical quantities such as density or concentration.

This expression can be viewed as an exact solution to a damped wave equation with a known source term. Specifically, consider the classical linear wave equation with exponential damping:
\begin{equation}
\frac{\partial^2 c}{\partial t^2} + 2\beta \frac{\partial c}{\partial t} = v^2 \nabla^2 c,
\label{eq:damped_wave_eq}
\end{equation}
where $v$ is the wave speed and $\beta$ is the damping coefficient. Looking for plane-wave solutions of the form
\[
c(x, y, t) = A \sin(k_x x + k_y y + \omega t + \theta_0) e^{-\beta t},
\]
and substituting into Eq.~\eqref{eq:damped_wave_eq} yields a dispersion relation
\[
\omega^2 - 2i \beta \omega = v^2 (k_x^2 + k_y^2),
\]
which is satisfied for appropriately chosen $\omega$ and $\beta$. 

The wave propagates along the direction of $\vec{k}$, with spatial periodicities $\lambda_x = 2\pi / k_x$ and $\lambda_y = 2\pi / k_y$. Temporally, it oscillates with angular frequency $\omega$ and decays exponentially at a rate governed by $\beta$. The random initial phase $\theta_0$ introduces sample diversity in the dataset.

With its analytical form and tunable parameters, the model offers a clean benchmark to evaluate the ability of learning models to capture structured spatiotemporal dynamics and assess generalization across resolutions.

\paragraph{Multi-Order-Parameter Grain Growth Model}

The second task investigates grain growth phenomena in polycrystalline materials, which is modeled using a multi-order-parameter phase-field framework~\cite{peng2023insights}. In this model, the microstructure is represented by a set of $N$ order parameters $\{ \eta_1(\mathbf{x}), \eta_2(\mathbf{x}), \dots, \eta_N(\mathbf{x}) \}$, where each $\eta_i$ corresponds to the presence of a specific grain orientation at spatial location $\mathbf{x}$. The evolution of the grain structure is driven by the minimization of the total free energy functional $F$, which is given by:
\begin{equation}
F[\{\eta_i\}] = \int_\Omega \left[ f(\eta_1, \eta_2, \dots, \eta_N) + \frac{\nu}{2} \sum_{i=1}^N |\nabla \eta_i|^2 \right] \, dV,
\label{eq:grain_free_energy}
\end{equation}
where $\nu$ is the gradient energy coefficient, and $f$ is the local homogeneous free energy density that governs the thermodynamic interaction among grain orientations.

The free energy density $f$ is constructed such that each order parameter $\eta_i$ favors values near $0$ or $1$, with strong penalization for overlapping grains. The expression reads:
\begin{equation}
f(\{\eta_i\}) = m \left[ \sum_{i=1}^N \left( \frac{\eta_i^4}{4} - \frac{\eta_i^2}{2} \right) + \frac{3}{2} \sum_{i=1}^N \sum_{j > i} \eta_i^2 \eta_j^2 + \frac{1}{4} \right],
\label{eq:grain_local_energy}
\end{equation}
where $m$ is a scaling constant. The first term in Eq.~\eqref{eq:grain_local_energy} ensures a double-well potential structure for each $\eta_i$, while the cross terms $\eta_i^2 \eta_j^2$ impose energetic penalties on spatial overlap between distinct grains, thus enforcing mutual exclusivity.

The temporal evolution of each order parameter $\eta_i$ follows the time-dependent Allen--Cahn equation, a second-order PDE driven by the variational derivative of the total free energy:
\begin{equation}
\frac{\partial \eta_i}{\partial t} = -L \frac{\delta F}{\delta \eta_i},
\label{eq:ac_grain}
\end{equation}
where $L$ is a positive kinetic coefficient controlling the rate of interface motion. Equation~\eqref{eq:ac_grain} leads to curvature-driven grain boundary migration and coarsening, allowing for the simulation of realistic microstructure evolution in materials without explicit tracking of grain interfaces.

This model forms a canonical benchmark for evaluating the ability of learning models to capture complex spatiotemporal dynamics in multi-field systems, including interface interaction, curvature effects, and topological transitions during grain coalescence and elimination.

\paragraph{Spinodal Decomposition via the Cahn--Hilliard Equation}

The third benchmark task focuses on spinodal decomposition, a spontaneous phase separation mechanism in binary mixtures. Unlike nucleation-driven processes, spinodal decomposition occurs within the unstable region of the free energy landscape, where infinitesimal fluctuations are sufficient to initiate phase separation. This phenomenon is governed by the \textit{Cahn--Hilliard} (C--H) equation, a fourth-order nonlinear PDE that captures domain formation and coarsening dynamics in the absence of nucleation.

The C--H equation models the time evolution of a conserved scalar field $c(\mathbf{x}, t)$, representing the molar fraction of one component in a binary alloy, as:
\begin{equation}
\frac{\partial c}{\partial t} = \nabla \cdot \left[ M(c) \nabla \left( \frac{\partial f_{\mathrm{chem}}}{\partial c} - \epsilon \nabla^2 c \right) \right],
\label{eq:ch}
\end{equation}
where $M(c) = M_0 c(1-c)$ is a concentration-dependent mobility ensuring zero flux at pure phases ($c=0,1$), and $\epsilon > 0$ is the gradient energy coefficient introducing interfacial energy regularization. 

The total free energy functional governing the system is given by:
\begin{equation}
F[c] = \int_\Omega \left( f_{\mathrm{chem}}(c) + \frac{\epsilon}{2} |\nabla c|^2 \right) d\mathbf{x},
\end{equation}
where $f_{\mathrm{chem}}(c)$ is the homogeneous chemical free energy density. In this task, we adopt the regular solution model:
\begin{equation}
f_{\mathrm{chem}}(c) = RT \left[ c \ln c + (1 - c) \ln(1 - c) \right] + \omega c(1 - c),
\label{eq:regular_solution}
\end{equation}
with $RT$ representing the thermal energy and $\omega > 0$ enforcing a miscibility gap that promotes phase separation.

By taking the variational derivative of $F[c]$ with respect to $c$, the chemical potential is obtained as:
\[
\mu = \frac{\delta F}{\delta c} = \frac{\partial f_{\mathrm{chem}}}{\partial c} - \epsilon \nabla^2 c,
\]
and substituting into the conserved dynamics yields the fourth-order equation~\eqref{eq:ch}.

The C--H equation provides a rich framework for modeling the evolution of interfacial patterns and long-time coarsening behavior in materials science. It serves as a canonical example to test the ability of machine learning models to capture high-order spatiotemporal dynamics with conservation constraints and nonlocal interactions.

\paragraph{Dendritic Solidification via the Kobayashi Phase-Field Model}

The fourth task models dendritic solidification using the classical anisotropic phase-field formulation. The system is described by a coupled temperature field $T$ and an order parameter $\varphi(\mathbf{x}, t)$, where $\varphi = 1$ denotes solid phase and $\varphi = 0$ corresponds to liquid. The total free energy functional governing the phase transition dynamics is given by:
\begin{equation}
F[\varphi, T] = \int_\Omega \left[ \frac{1}{2} \epsilon(\theta)^2 |\nabla \varphi|^2 + f(\varphi, T) \right] \, d\mathbf{x},
\label{eq:dendrite_energy}
\end{equation}
where $\epsilon(\theta)$ is an orientation-dependent anisotropic gradient coefficient and $f(\varphi, T)$ is the bulk free energy density modeled by a double-well potential.

Anisotropy is introduced through the orientation-dependent coefficient
\[
\epsilon(\theta) = \epsilon_0 \left[ 1 + \delta \cos(n(\theta - \theta_0)) \right], \quad \theta = \arctan\left( -\frac{\varphi_y}{\varphi_x} \right),
\]
where $\epsilon_0$ is a baseline surface energy, $\delta$ controls the anisotropy strength, $n$ determines the symmetry (e.g., $n=4$ for fourfold), and $\theta_0$ is the reference crystal orientation.

The chemical bulk free energy density $f(\varphi, T)$ is given by:
\begin{equation}
f(\varphi, T) = \frac{1}{4} \varphi^4 - \left[ \frac{1}{2} - \frac{1}{3} m(T) \right] \varphi^3 + \left[ \frac{1}{4} - \frac{1}{2} m(T) \right] \varphi^2,
\label{eq:dendrite_bulk}
\end{equation}
where $m(T)$ is a temperature-dependent function defined by
\[
m(T) = \frac{\alpha}{\pi} \arctan \left[ \gamma (T_{\text{eq}} - T) \right],
\]
with $\alpha$ and $\gamma$ controlling the magnitude and sharpness of the phase transition, and $T_{\text{eq}}$ being the equilibrium temperature at which the solid and liquid phases coexist.

The system evolves via a set of coupled partial differential equations. The phase-field equation is a time-dependent Allen--Cahn-type equation:
\begin{equation}
\tau \frac{\partial \varphi}{\partial t} = - \frac{\delta F}{\delta \varphi},
\label{eq:dendrite_phi}
\end{equation}
and the temperature field evolves according to a heat diffusion equation with latent heat feedback:
\begin{equation}
\frac{\partial T}{\partial t} = \nabla^2 T + K \frac{\partial \varphi}{\partial t},
\label{eq:dendrite_temp}
\end{equation}
where $\tau$ is the phase-field relaxation time and $K$ is the latent heat parameter coupling thermal diffusion with phase change.

This coupled system captures key aspects of dendritic solidification including anisotropic interfacial energy, temperature-dependent solidification front dynamics, and latent heat release. As such, it provides a high-fidelity benchmark for testing the ability of learning models to resolve strongly coupled, nonlinear, and anisotropic spatiotemporal processes.

\textbf{\textit{As part of our ongoing commitment to expanding the benchmark for a more holistic evaluation, we are now introducing an additional dataset.}}

\paragraph{Dumbbell Evolution via the Allen--Cahn Phase-Field Model}
The additional benchmark examines curvature-driven geometric evolution using the Allen--Cahn phase-field formulation. The system models the interface motion under mean curvature flow, where the normal velocity equals the local geometric curvature~\cite{deckelnick2005computation}. A smooth order parameter $u(\mathbf{x}, t)\in[0,1]$ distinguishes the two phases, and the interfacial energy is represented by the diffuse-interface functional
\begin{equation}
P_\varepsilon(u)
=
\int_{\mathbb{R}^d}
\left[
\varepsilon \frac{|\nabla u|^2}{2}
+
\frac{1}{\varepsilon} W(u)
\right]
\, d\mathbf{x},
\label{eq:dumbbell_energy}
\end{equation}
where $\varepsilon > 0$ controls the interfacial thickness and $W(u)=\tfrac12 u^2 (1-u)^2$ is a symmetric double-well potential.

Taking the $L^2$--gradient flow of $P_\varepsilon$ yields the classical Allen--Cahn equation
\begin{equation}
\frac{\partial u}{\partial t}
=
\Delta u
-
\frac{1}{\varepsilon^2} W'(u),
\label{eq:dumbbell_ac}
\end{equation}
which drives the diffuse interface to move with velocity proportional to its mean curvature. In the sharp-interface limit $\varepsilon\to 0$, the zero level set of $u$ evolves according to mean curvature flow, with normal velocity $V_n = H$.

For this Dumbbell task, the initial condition is generated from the signed distance function to a two-lobed dumbbell-shaped domain $\Omega(0)$:
\begin{equation}
u_\varepsilon(\mathbf{x}, 0)
=
q\!\left( \frac{d(\mathbf{x}, \Omega(0))}{\varepsilon} \right),~~
q(s)=\tfrac12\left( 1 - \tanh\!\left(\frac{s}{2}\right) \right),
\label{eq:dumbbell_init}
\end{equation}
where $d(\mathbf{x},\Omega(0))$ denotes the signed distance. The function $q$ represents the optimal 1D transition profile minimizing the interfacial energy~\cite{bretin2022learning}.

During evolution, curvature-driven motion causes the narrow neck connecting the two lobes to progressively thin and ultimately pinch off, producing two separate droplets. This process captures geometric singularity formation (topological transition) and curvature-induced interface dynamics within the phase-field paradigm, making it a compelling extended example for our benchmark.

\begin{sidewaystable*}[htbp]

  \centering

  \caption{More Details of the Main Results.}
    \resizebox{1\linewidth}{!}{\begin{tabular}{cccccccccccccccccccccc}
    \toprule
          &       & \multicolumn{5}{c}{Plane-wave propagation} & \multicolumn{5}{c}{Grain growth}      & \multicolumn{5}{c}{Spinodal decomposition} & \multicolumn{5}{c}{Dendritic solidification} \\
    \midrule
    Model domain & Model & MAE   & MSE   & RMSE  & SSIM  & L-ETAP & MAE   & MSE   & RMSE  & SSIM  & L-EAPSR & MAE   & MSE   & RMSE  & SSIM  & L-ETAP & MAE   & MSE   & RMSE  & SSIM  & L-ETAP \\
    \midrule
    \multicolumn{22}{c}{Short term prediction (10-10)} \\
    \midrule
    \multicolumn{1}{c}{\multirow{5}[2]{*}{Microstructural}} & E3DLSTM & 0.00732 & 0.00028 & 0.01673 & 0.99315 & -0.993 & 0.01018 & 0.00115 & 0.03394 & 0.98607 & -2.488 & 0.00069 & 0.00002 & 0.00418 & 0.99970 & -2.826 & 0.00263 & 0.00016 & 0.01275 & 0.99717 & -2.540 \\
          & ConvGRU & 0.00278 & 0.00001 & 0.00338 & 0.99944 & -0.994 & 0.00769 & 0.00046 & 0.02134 & 0.99183 & -2.586 & 0.00063 & 0.00001 & 0.00389 & 0.99967 & -2.853 & 0.00097 & 0.00001 & 0.00267 & 0.99965 & -2.865 \\
          & PredRNN & 0.00062 & 0.00000 & 0.00102 & 0.99994 & -0.995 & 0.00703 & 0.00060 & 0.02453 & 0.99165 & -2.541 & 0.00047 & \textbf{0.00001} & \textbf{0.00270} & 0.99985 & \textbf{-2.916} & 0.00145 & 0.00003 & 0.00574 & 0.99911 & -2.737 \\
          & ConvLSTM & 0.00216 & 0.00001 & 0.00328 & 0.99965 & \textbf{-0.996} & 0.01073 & 0.00095 & 0.03082 & 0.98707 & -2.492 & \textbf{0.00002} & 0.00002 & 0.00423 & 0.99969 & -2.857 & 0.00143 & 0.00003 & 0.00545 & 0.99912 & -2.726 \\
          & VMamba & 0.00083 & 0.00000 & 0.00151 & 0.99991 & -0.993 & \textbf{0.00226} & \textbf{0.00008} & \textbf{0.00871} & \textbf{0.99926} & \textbf{-2.875} & 0.00055 & 0.00001 & 0.00382 & 0.99983 & -2.902 & \textbf{0.00056} & \textbf{0.00000} & \textbf{0.00259} & \textbf{0.99990} & \textbf{-2.942} \\
    \midrule
    \multicolumn{1}{c}{\multirow{9}[2]{*}{Spatiotemporal}} & PredRNN++ & 0.00068 & 0.00000 & 0.00108 & 0.99993 & -0.993 & 0.01003 & 0.00094 & 0.03071 & 0.98795 & -2.545 & 0.00050 & 0.00001 & 0.00285 & 0.99981 & -2.844 & 0.00171 & 0.00004 & 0.00662 & 0.99883 & -2.702 \\
          & MAU   & 0.00550 & 0.00008 & 0.00867 & 0.99732 & -0.977 & 0.02159 & 0.00353 & 0.05939 & 0.95131 & -1.830 & 0.00197 & 0.00006 & 0.00773 & 0.99876 & -2.494 & 0.00210 & 0.00006 & 0.00752 & 0.99828 & -2.553 \\
          & SimVP & 0.00185 & 0.00002 & 0.00471 & 0.99905 & -0.993 & 0.00936 & 0.00066 & 0.02563 & 0.99139 & -2.600 & 0.00097 & 0.00002 & 0.00492 & 0.99961 & -2.678 & 0.00163 & 0.00004 & 0.00646 & 0.99887 & -2.665 \\
          & PredRNN.v2 & 0.00100 & 0.00000 & 0.00157 & 0.99986 & -0.993 & 0.01400 & 0.00156 & 0.03954 & 0.97898 & -2.421 & 0.00081 & 0.00002 & 0.00425 & 0.99964 & -2.844 & 0.00137 & 0.00003 & 0.00561 & 0.99915 & -2.739 \\
          & TAU   & 0.00168 & 0.00001 & 0.00271 & 0.99942 & -0.992 & 0.00567 & 0.00045 & 0.02127 & 0.99463 & -2.633 & 0.00054 & 0.00002 & 0.00421 & 0.99973 & -2.864 & 0.00103 & 0.00002 & 0.00441 & 0.99954 & -2.741 \\
          & SwinLSTM & 0.00165 & 0.00001 & 0.00236 & 0.99938 & -0.995 & 0.01151 & 0.00107 & 0.03269 & 0.98502 & -2.570 & 0.00113 & 0.00001 & 0.00354 & 0.99932 & -2.868 & 0.00171 & 0.00006 & 0.00764 & 0.99877 & -2.612 \\
          & VRMNN & \textbf{0.00053} & \textbf{0.00000} & \textbf{0.00081} & \textbf{0.99998} & -0.995 & 0.00821 & 0.00050 & 0.02230 & 0.99602 & -2.618 & 0.00101 & 0.00001 & 0.00347 & \textbf{0.99987} & -2.843 & 0.00153 & 0.00004 & 0.00668 & 0.99962 & -2.673 \\
          & SimVP.v2 & 0.00356 & 0.00003 & 0.00578 & 0.99920 & -0.993 & 0.00619 & 0.00056 & 0.02366 & 0.99339 & -2.603 & 0.00092 & 0.00002 & 0.00442 & 0.99954 & -2.755 & 0.00115 & 0.00002 & 0.00495 & 0.99945 & -2.638 \\
          & PredFormer & 0.00698 & 0.00008 & 0.00917 & 0.98603 & -0.972 & 0.00703 & 0.00060 & 0.02453 & 0.99165 & -2.696 & 0.00732 & 0.00016 & 0.01273 & 0.96767 & -2.243 & 0.00542 & 0.00016 & 0.01261 & 0.95548 & -2.443 \\
    \midrule
    \multicolumn{22}{c}{Long term prediction (10-50)} \\
    \midrule
    \multicolumn{1}{c}{\multirow{5}[2]{*}{Microstructural}} & E3DLSTM & 0.06741 & 0.01439 & 0.11997 & 0.78022 & -0.667 & 0.03814 & 0.01224 & 0.11065 & 0.87022 & -1.750 & \textbf{0.00038} & 0.00370 & 0.01951 & 0.99546 & -1.809 & 0.00955 & 0.00252 & 0.05015 & 0.97352 & -1.188 \\
          & ConvGRU & 0.01402 & 0.00034 & 0.01845 & 0.98973 & -0.714 & 0.02849 & 0.00782 & 0.08843 & 0.91765 & -1.760 & 0.00255 & 0.00028 & 0.01691 & 0.99727 & -2.139 & 0.00474 & 0.00071 & 0.02720 & 0.99092 & -1.708 \\
          & PredRNN & 0.00480 & 0.00007 & 0.00819 & 0.99742 & \textbf{-0.730} & 0.13631 & 0.04916 & 0.22172 & 0.46442 & -0.685 & 0.00290 & 0.00031 & 0.01758 & 0.99655 & -1.817 & 0.09786 & 0.06716 & 0.25915 & 0.71776 & -0.032 \\
          & ConvLSTM & 0.07174 & 0.01362 & 0.11670 & 0.73850 & -0.622 & 0.03433 & 0.01005 & 0.10027 & 0.89229 & -1.564 & 0.00477 & 0.00064 & 0.02520 & 0.99230 & -1.783 & 0.00776 & 0.00172 & 0.04143 & 0.98008 & -1.278 \\
          & VMamba & \textbf{0.00380} & 0.00007 & 0.00817 & 0.99709 & -0.728 & \textbf{0.00633} & \textbf{0.00089} & \textbf{0.02986} & \textbf{0.99008} & \textbf{-2.283} & 0.00198 & \textbf{0.00018} & \textbf{0.01342} & \textbf{0.99848} & \textbf{-2.221} & \textbf{0.00249} & \textbf{0.00022} & \textbf{0.01922} & \textbf{0.99609} & \textbf{-1.983} \\
    \midrule
    \multicolumn{1}{c}{\multirow{9}[2]{*}{Spatiotemporal}} & PredRNN++ & 0.00447 & \textbf{0.00006} & \textbf{0.00754} & \textbf{0.99803} & -0.704 & 0.03330 & 0.00985 & 0.09926 & 0.89755 & -1.764 & 0.00396 & 0.00053 & 0.02294 & 0.99393 & -1.916 & 0.00833 & 0.00184 & 0.04285 & 0.97866 & -1.320 \\
          & MAU   & 0.09679 & 0.02507 & 0.15833 & 0.65653 & -0.411 & 0.07337 & 0.02789 & 0.16699 & 0.68662 & -1.093 & 0.01314 & 0.00256 & 0.05056 & 0.96809 & -1.144 & 0.02225 & 0.00917 & 0.09578 & 0.92582 & -0.813 \\
          & SimVP & 0.00829 & 0.00042 & 0.02054 & 0.98538 & -0.725 & 0.03356 & 0.00890 & 0.09435 & 0.89925 & -1.899 & 0.00359 & 0.00052 & 0.02279 & 0.99565 & -1.820 & 0.00573 & 0.00104 & 0.03226 & 0.98725 & -1.777 \\
          & PredRNN.v2 & 0.00658 & 0.00013 & 0.01160 & 0.99545 & -0.708 & 0.09531 & 0.02865 & 0.16927 & 0.48566 & -0.487 & 0.00478 & 0.00062 & 0.02493 & 0.99300 & -1.510 & 0.00752 & 0.00161 & 0.04015 & 0.98058 & -1.402 \\
          & TAU   & 0.00808 & 0.00019 & 0.01396 & 0.99389 & -0.727 & 0.02466 & 0.00681 & 0.08250 & 0.92942 & -1.868 & 0.00180 & 0.00027 & 0.01643 & 0.99809 & -2.213 & 0.00402 & 0.00056 & 0.02374 & 0.99233 & -1.797 \\
          & SwinLSTM & 0.01194 & 0.00042 & 0.02046 & 0.98421 & -0.726 & 0.04791 & 0.01594 & 0.12624 & 0.81917 & -1.805 & 0.00340 & 0.00030 & 0.01745 & 0.99635 & -2.000 & 0.00756 & 0.00188 & 0.04339 & 0.98029 & -1.513 \\
          & VRMNN & 0.00617 & 0.00012 & 0.01113 & 0.99683 & -0.728 & 0.01982 & 0.00390 & 0.06248 & 0.96615 & -1.975 & 0.00347 & 0.00036 & 0.01888 & 0.99753 & -1.911 & 0.00599 & 0.00124 & 0.03521 & 0.98984 & -1.651 \\
          & SimVP.v2 & 0.01329 & 0.00050 & 0.02234 & 0.98652 & -0.724 & 0.02775 & 0.00798 & 0.08931 & 0.91645 & -1.851 & 0.00293 & 0.00038 & 0.01950 & 0.99679 & -2.063 & 0.00441 & 0.00068 & 0.02615 & 0.99109 & -1.762 \\
          & PredFormer & 0.02008 & 0.00084 & 0.02892 & 0.95577 & -0.695 & 0.02798 & 0.00714 & 0.08450 & 0.91212 & -2.022 & 0.01852 & 0.00311 & 0.05575 & 0.93378 & -1.332 & 0.01411 & 0.00344 & 0.05868 & 0.92531 & -1.380 \\
    \midrule
    \multicolumn{22}{c}{Long term prediction (10-90)} \\
    \midrule
    \multicolumn{1}{c}{\multirow{5}[2]{*}{Microstructural}} & E3DLSTM & 0.11511 & 0.02920 & 0.17089 & 0.58738 & -0.489 & 0.05756 & 0.02191 & 0.14800 & 0.76496 & -1.450 & 0.00769 & 0.00133 & 0.03642 & 0.98552 & -1.441 & 0.04798 & 0.02718 & 0.16485 & 0.84943 & -0.212 \\
          & ConvGRU & 0.02794 & 0.00132 & 0.03635 & 0.95884 & -0.596 & 0.04593 & 0.01611 & 0.12688 & 0.82782 & -1.425 & 0.00482 & 0.00080 & 0.02791 & 0.99304 & -1.843 & 0.01798 & 0.00675 & 0.08216 & 0.95175 & -1.070 \\
          & PredRNN & 0.01583 & 0.00069 & 0.02630 & 0.97109 & -0.598 & 0.22240 & 0.10165 & 0.31882 & 0.26338 & 0.060 & 0.00764 & 0.00140 & 0.03744 & 0.98465 & -1.301 & 0.23598 & 0.18244 & 0.42713 & 0.43134 & 0.296 \\
          & ConvLSTM & 0.11438 & 0.02592 & 0.16098 & 0.54013 & -0.442 & 0.05234 & 0.01890 & 0.13749 & 0.79176 & -1.173 & 0.01282 & 0.00332 & 0.05765 & 0.96745 & -1.268 & 0.02984 & 0.01472 & 0.12132 & 0.90480 & -0.489 \\
          & VMamba & \textbf{0.00804} & \textbf{0.00027} & \textbf{0.01652} & \textbf{0.98745} & -0.629 & \textbf{0.01061} & \textbf{0.00223} & \textbf{0.04727} & \textbf{0.97322} & \textbf{-2.011} & 0.00352 & \textbf{0.00047} & \textbf{0.02138} & \textbf{0.99627} & -1.944 & \textbf{0.00876} & \textbf{0.00307} & \textbf{0.05542} & \textbf{0.97682} & \textbf{-1.437} \\
    \midrule
    \multicolumn{1}{c}{\multirow{9}[2]{*}{Spatiotemporal}} & PredRNN++ & 0.01389 & 0.00052 & 0.02281 & 0.98039 & \textbf{-0.633} & 0.05055 & 0.01821 & 0.13494 & 0.80531 & -1.458 & 0.01090 & 0.00271 & 0.05207 & 0.97385 & -1.437 & 0.03135 & 0.01573 & 0.12542 & 0.90127 & -0.485 \\
          & MAU   & 0.12871 & 0.03413 & 0.18475 & 0.51672 & -0.340 & 0.09849 & 0.04144 & 0.20358 & 0.53944 & -0.871 & 0.02655 & 0.00892 & 0.09445 & 0.92043 & -0.760 & 0.05532 & 0.03457 & 0.18592 & 0.82421 & -0.349 \\
          & SimVP & 0.01654 & 0.00127 & 0.03558 & 0.95929 & -0.621 & 0.05234 & 0.01795 & 0.13397 & 0.80088 & -1.605 & 0.00629 & 0.00128 & 0.03582 & 0.98976 & -1.556 & 0.01487 & 0.00536 & 0.07325 & 0.95754 & -1.179 \\
          & PredRNN.v2 & 0.02097 & 0.00121 & 0.03477 & 0.96081 & \textbf{-0.633} & 0.13292 & 0.04068 & 0.20170 & 0.28880 & -0.492 & 0.01341 & 0.00325 & 0.05702 & 0.96639 & -0.888 & 0.02810 & 0.01380 & 0.11749 & 0.91178 & -0.587 \\
          & TAU   & 0.01585 & 0.00069 & 0.02623 & 0.97736 & -0.626 & 0.04177 & 0.01464 & 0.12098 & 0.84532 & -1.557 & \textbf{0.00310} & 0.00063 & 0.02517 & 0.99590 & \textbf{-1.958} & 0.01140 & 0.00352 & 0.05931 & 0.96924 & -1.209 \\
          & SwinLSTM & 0.03168 & 0.00277 & 0.05263 & 0.91687 & -0.594 & 0.07515 & 0.03012 & 0.17355 & 0.66420 & -1.408 & 0.00651 & 0.00105 & 0.03240 & 0.98954 & -1.620 & 0.00957 & 0.02076 & 0.09784 & 0.93640 & -0.956 \\
          & VRMNN & 0.02028 & 0.00124 & 0.03528 & 0.96877 & -0.615 & 0.03013 & 0.00826 & 0.09091 & 0.91929 & -1.697 & 0.00640 & 0.00103 & 0.03202 & 0.99288 & -1.524 & 0.01915 & 0.00847 & 0.09203 & 0.94758 & -1.140 \\
          & SimVP.v2 & 0.02404 & 0.00150 & 0.03878 & 0.95755 & -0.617 & 0.04612 & 0.01655 & 0.12863 & 0.82407 & -1.536 & 0.00506 & 0.00090 & 0.03004 & 0.99268 & -1.811 & 0.01217 & 0.00398 & 0.06307 & 0.96618 & -1.149 \\
          & PredFormer & 0.04017 & 0.00340 & 0.05832 & 0.87576 & -0.560 & 0.04446 & 0.01460 & 0.12084 & 0.82722 & -1.741 & 0.03083 & 0.00919 & 0.09589 & 0.89108 & -0.990 & 0.03474 & 0.01697 & 0.13028 & 0.85560 & -0.823 \\
    \bottomrule
    \end{tabular}}
  \label{tab:More Details of the Main Results}%
\end{sidewaystable*}%

\begin{sidewaystable*}[htbp]

  \centering

  \caption{The Variance of the Main Results.}
    \resizebox{1\linewidth}{!}{\begin{tabular}{cccccccccccccccccc}
    \toprule
          &       & \multicolumn{4}{c}{Plane-wave propagation} & \multicolumn{4}{c}{Grain growth}      & \multicolumn{4}{c}{Spinodal decomposition} & \multicolumn{4}{c}{Dendritic solidification} \\
    \midrule
    Model domain & Model & Var(MAE)   & Var(MSE)   & Var(RMSE)  & Var(SSIM)   & Var(MAE)   & Var(MSE)   & Var(RMSE)  & Var(SSIM)   & Var(MAE)   & Var(MSE)   & Var(RMSE)  & Var(SSIM)   & Var(MAE)   & Var(MSE)   & Var(RMSE)  & Var(SSIM)  \\
    \midrule
    \multicolumn{18}{c}{Short term prediction (10-10)} \\
    \midrule
    \multicolumn{1}{c}{\multirow{5}[2]{*}{Microstructural}} & E3DLSTM & 0.000082 & 0.000001 & 0.000066 & 0.000077 & 0.000013 & 0.000001 & 0.000087 & 0.000021 &0.000000 &0.000000 & 0.000013 &0.000000 &0.000000 &0.000000 & 0.000003 &0.000000 \\
  & ConvGRU & 0.000014 &0.000000 & 0.000021 & 0.000001 & 0.000002 & 0.000000 & 0.000044 & 0.000004 &0.000000 &0.000000 & 0.000008 &0.000000 & 0.000000 & 0.000000 & 0.000001 & 0.000000 \\
  & PredRNN & 0.000002 & 0.000000 & 0.000004 &0.000000 & 0.000014 & 0.000001 & 0.000092 & 0.000013 & 0.000000 & 0.000000 & 0.000007 &0.000000 & 0.000001 &0.000000 & 0.000008 &0.000000 \\
  & ConvLSTM & 0.000032 &0.000000 & 0.000023 &0.000000 & 0.000017 & 0.000001 & 0.000091 & 0.000021 &0.000000 &0.000000 & 0.000013 &0.000000 &0.000000 &0.000000 & 0.000001 & 0.000000 \\
  & VMamba & 0.000004 &0.000000 & 0.000006 &0.000000 & 0.000001 &0.000000 & 0.000026 & 0.000002 &0.000000 &0.000000 & 0.000011 &0.000000 &0.000000 &0.000000 & 0.000001 &0.000000 \\

    \midrule
    \multicolumn{1}{c}{\multirow{9}[2]{*}{Spatiotemporal}} & PredRNN++ & 0.000003 &0.000000 & 0.000004 &0.000000 & 0.000015 & 0.000001 & 0.000088 & 0.000019 &0.000000 &0.000000 & 0.000009 &0.000000 &0.000000 &0.000000 & 0.000002 & 0.000000 \\
  & MAU & 0.000040 & 0.000001 & 0.000089 & 0.000132 & 0.000024 & 0.000002 & 0.000083 & 0.000055 &0.000000 &0.000000 & 0.000033 & 0.000003 &0.000000 &0.000000 & 0.000002 &0.000000 \\
  & SimVP & 0.000060 & 0.000001 & 0.000086 & 0.000091 & 0.000010 & 0.000001 & 0.000077 & 0.000001 &0.000000 &0.000000 & 0.000014 & 0.000001 &0.000000 &0.000000 & 0.000002 & 0.000000 \\
  & PredRNN.v2 & 0.000001 &0.000000 & 0.000002 &0.000000 & 0.000023 & 0.000001 & 0.000111 & 0.000038 &0.000000 &0.000000 & 0.000015 & 0.000001 &0.000000 &0.000000 & 0.000001 &0.000000 \\
  & TAU & 0.000012 &0.000000 & 0.000016 & 0.000002 & 0.000005 &0.000000 & 0.000080 & 0.000009 &0.000000 &0.000000 & 0.000019 & 0.000001 &0.000000 &0.000000 & 0.000002 &0.000000 \\
  & SwinLSTM & 0.000001 &0.000000 & 0.000022 & 0.000009 & 0.000022 & 0.000001 & 0.000122 & 0.000053 & 0.000001 &0.000000 & 0.000017 & 0.000003 &0.000000 &0.000000 & 0.000002 &0.000000 \\
  & VMRNN & 0.000000 & 0.000000 & 0.000000 & 0.000000 & 0.000008 & 0.000000 & 0.000062 & 0.000010 & 0.000000 & 0.000000 & 0.000008 & 0.000000 & 0.000000 & 0.000000 & 0.000001 & 0.000000 \\
  & SimVP.v2 & 0.000047 &0.000000 & 0.000062 & 0.000023 & 0.000008 &0.000000 & 0.000099 & 0.000013 &0.000000 &0.000000 & 0.000012 & 0.000001 & 0.000001 &0.000000 & 0.000007 &0.000000 \\
  & PredFormer & 0.000049 & 0.000000 & 0.000068 & 0.000089 & 0.000014 & 0.000001 & 0.000115 & 0.000036 & 0.000001 & 0.000000 & 0.000027 & 0.000015 & 0.000000 & 0.000000 & 0.000013 & 0.000008 \\

    \midrule
    \multicolumn{18}{c}{Long term prediction (10-50)} \\
    \midrule
    \multicolumn{1}{c}{\multirow{5}[2]{*}{Microstructural}} & E3DLSTM & 0.000330 & 0.000016 & 0.000688 & 0.002110 & 0.000060 & 0.000014 & 0.000207 & 0.001009 & 0.000002 & 0.000001 & 0.000145 & 0.000019 & 0.000005 & 0.000002 & 0.000156 & 0.000190 \\
  & ConvGRU & 0.000316 & 0.000005 & 0.000519 & 0.000818 & 0.000028 & 0.000007 & 0.000165 & 0.000444 & 0.000003 & 0.000001 & 0.000162 & 0.000026 & 0.000005 & 0.000001 & 0.000179 & 0.000120 \\
  & PredRNN & 0.000062 &0.000000 & 0.000133 & 0.000120 & 0.000028 & 0.000017 & 0.000068 & 0.000144 & 0.000002 & 0.000001 & 0.000123 & 0.000017 & 0.000200 & 0.000133 & 0.000456 & 0.000476 \\
  & ConvLSTM & 0.000993 & 0.000066 & 0.001356 & 0.001690 & 0.000110 & 0.000021 & 0.000349 & 0.001380 & 0.000020 & 0.000002 & 0.000023 & 0.000025 & 0.000011 & 0.000003 & 0.000319 & 0.000303 \\
  & VMamba & 0.000074 &0.000000 & 0.000125 & 0.000149 & 0.000018 & 0.000003 & 0.000357 & 0.000295 & 0.000002 & 0.000001 & 0.000107 & 0.000019 & 0.000001 &0.000000 & 0.000038 & 0.000006 \\

    \midrule
    \multicolumn{1}{c}{\multirow{9}[2]{*}{Spatiotemporal}} & PredRNN++ & 0.000055 &0.000000 & 0.000110 & 0.000124 & 0.000102 & 0.000019 & 0.000333 & 0.001253 & 0.000004 & 0.000001 & 0.000134 & 0.000031 & 0.000003 &0.000000 & 0.000100 & 0.000050 \\
  & MAU & 0.000412 & 0.000019 & 0.000912 & 0.001195 & 0.000075 & 0.000020 & 0.000137 & 0.001513 & 0.000013 & 0.000005 & 0.000235 & 0.000162 & 0.000028 & 0.000015 & 0.000454 & 0.000903 \\
  & SimVP & 0.000689 & 0.000045 & 0.001064 & 0.005579 & 0.000085 & 0.000015 & 0.000312 & 0.001191 & 0.000009 & 0.000001 & 0.000099 & 0.000020 & 0.000006 & 0.000001 & 0.000227 & 0.000153 \\
  & PredRNN.v2 & 0.000028 &0.000000 & 0.000057 & 0.000077 & 0.000055 & 0.000014 & 0.000294 & 0.000986 & 0.000005 & 0.000002 & 0.000214 & 0.000049 & 0.000018 & 0.000007 & 0.000483 & 0.000535 \\
  & TAU & 0.000191 & 0.000004 & 0.000308 & 0.000808 & 0.000071 & 0.000014 & 0.000333 & 0.000977 & 0.000004 & 0.000001 & 0.000208 & 0.000036 & 0.000003 &0.000000 & 0.000119 & 0.000067 \\
  & SwinLSTM & 0.000399 & 0.000015 & 0.000614 & 0.003368 & 0.000141 & 0.000029 & 0.000317 & 0.001997 & 0.000003 & 0.000001 & 0.000199 & 0.000069 & 0.000010 & 0.000002 & 0.000401 & 0.000228 \\
  & VRMNN & 0.000042 & 0.000000 & 0.000098 & 0.000114 & 0.000052 & 0.000009 & 0.000357 & 0.000700 & 0.000004 & 0.000002 & 0.000187 & 0.000043 & 0.000005 & 0.000001 & 0.000217 & 0.000086 \\
  & SimVP.v2 & 0.000458 & 0.000020 & 0.000702 & 0.002441 & 0.000085 & 0.000016 & 0.000341 & 0.001058 & 0.000004 & 0.000001 & 0.000185 & 0.000031 & 0.000007 &0.000000 & 0.000151 & 0.000040 \\
  & PredFormer & 0.000392 & 0.000011 & 0.000571 & 0.001709 & 0.000106 & 0.000020 & 0.000448 & 0.001678 & 0.000016 & 0.000005 & 0.000258 & 0.000169 & 0.000031 & 0.000013 & 0.000753 & 0.000813 \\

    \midrule
    \multicolumn{18}{c}{Long term prediction (10-90)} \\
    \midrule
    \multicolumn{1}{c}{\multirow{5}[2]{*}{Microstructural}} & E3DLSTM & 0.000993 & 0.000113 & 0.000121 & 0.017460 & 0.000077 & 0.000022 & 0.000197 & 0.002259 & 0.000012 & 0.000005 & 0.000391 & 0.000135 & 0.000175 & 0.000098 & 0.000888 & 0.001525 \\
  & ConvGRU & 0.000934 & 0.000045 & 0.001557 & 0.010206 & 0.000054 & 0.000016 & 0.000191 & 0.001389 & 0.000011 & 0.000005 & 0.000357 & 0.000098 & 0.000072 & 0.000026 & 0.000798 & 0.001251 \\
  & PredRNN & 0.000224 & 0.000004 & 0.000422 & 0.002296 & 0.000045 & 0.000050 & 0.000097 & 0.000041 & 0.000010 & 0.000004 & 0.000291 & 0.000105 & 0.000348 & 0.000318 & 0.000515 & 0.001471 \\
  & ConvLSTM & 0.001065 & 0.000096 & 0.002163 & 0.021800 & 0.000158 & 0.000038 & 0.000384 & 0.003488 & 0.000022 & 0.000009 & 0.000384 & 0.000158 & 0.000257 & 0.000127 & 0.001931 & 0.004031 \\
  & VMamba & 0.000221 & 0.000003 & 0.000366 & 0.001609 & 0.000062 & 0.000014 & 0.000816 & 0.001460 & 0.000008 & 0.000003 & 0.000241 & 0.000074 & 0.000018 & 0.000004 & 0.000322 & 0.000219 \\

    \midrule
    \multicolumn{1}{c}{\multirow{9}[2]{*}{Spatiotemporal}} & PredRNN++ & 0.000341 & 0.000009 & 0.000556 & 0.001898 & 0.000155 & 0.000037 & 0.000392 & 0.003358 & 0.000015 & 0.000004 & 0.000282 & 0.000144 & 0.000035 & 0.000013 & 0.000427 & 0.000653 \\
  & MAU & 0.001692 & 0.000126 & 0.001923 & 0.013549 & 0.000069 & 0.000023 & 0.000110 & 0.001765 & 0.000045 & 0.000022 & 0.000497 & 0.000726 & 0.000134 & 0.000106 & 0.000680 & 0.001725 \\
  & SimVP & 0.001288 & 0.000088 & 0.001951 & 0.019977 & 0.000018 & 0.000041 & 0.000432 & 0.003019 & 0.000011 & 0.000004 & 0.000331 & 0.000076 & 0.000076 & 0.000023 & 0.001056 & 0.001282 \\
  & PredRNN.v2 & 0.000139 & 0.000004 & 0.000276 & 0.002563 & 0.000099 & 0.000023 & 0.000392 & 0.002484 & 0.000016 & 0.000010 & 0.000308 & 0.000103 & 0.000282 & 0.000155 & 0.001704 & 0.003899 \\
  & TAU & 0.000483 & 0.000019 & 0.000751 & 0.005430 & 0.000111 & 0.000028 & 0.000355 & 0.002671 & 0.000012 & 0.000005 & 0.000417 & 0.000116 & 0.000038 & 0.000009 & 0.000587 & 0.000558 \\
  & SwinLSTM & 0.000934 & 0.000066 & 0.001661 & 0.010421 & 0.000157 & 0.000038 & 0.000249 & 0.002839 & 0.000009 & 0.000004 & 0.000233 & 0.000446 & 0.000166 & 0.000067 & 0.002043 & 0.002812 \\
  & VRMNN & 0.000409 & 0.000018 & 0.000901 & 0.005297 & 0.000104 & 0.000025 & 0.000551 & 0.002479 & 0.000015 & 0.000006 & 0.000393 & 0.000152 & 0.000095 & 0.000035 & 0.001263 & 0.001480 \\
  & SimVP.v2 & 0.000947 & 0.000063 & 0.001427 & 0.011695 & 0.000123 & 0.000030 & 0.000338 & 0.002850 & 0.000012 & 0.000005 & 0.000397 & 0.000110 & 0.000051 & 0.000012 & 0.000764 & 0.000810 \\
  & PredFormer & 0.000979 & 0.000072 & 0.001565 & 0.011022 & 0.000145 & 0.000036 & 0.000467 & 0.003466 & 0.000048 & 0.000023 & 0.000488 & 0.000522 & 0.000251 & 0.000136 & 0.002075 & 0.004706 \\

    \bottomrule
    \end{tabular}}
  \label{tab:Cross-Sample Variance of the Main Results}%
\end{sidewaystable*}%

\begin{figure*}[ht]
  \centering
  \includegraphics[width=1\linewidth]{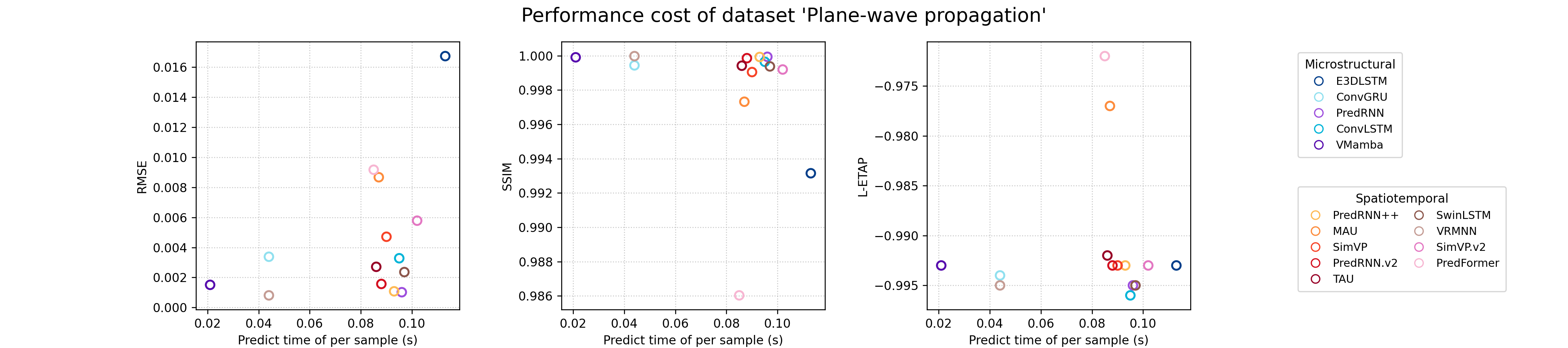}
  \caption{The performance-cost of Plane-wave propagation.}
  \label{fig_perf_plane}
\end{figure*}

\begin{figure*}[ht]
  \centering
  \includegraphics[width=1\linewidth]{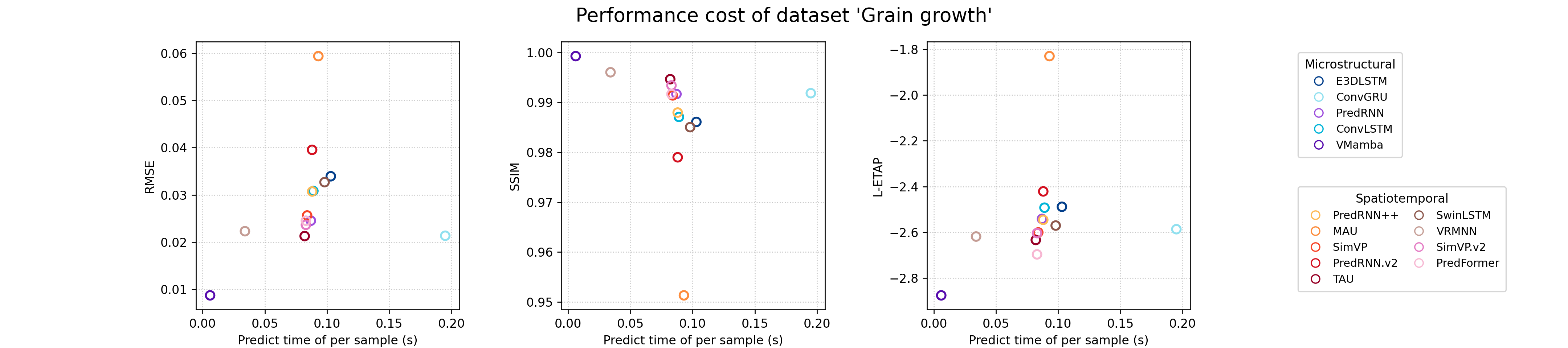}
  \caption{The performance-cost of Grain growth.}
  \label{fig_perf_grain}
\end{figure*}

\begin{figure*}[ht]
  \centering
  \includegraphics[width=1\linewidth]{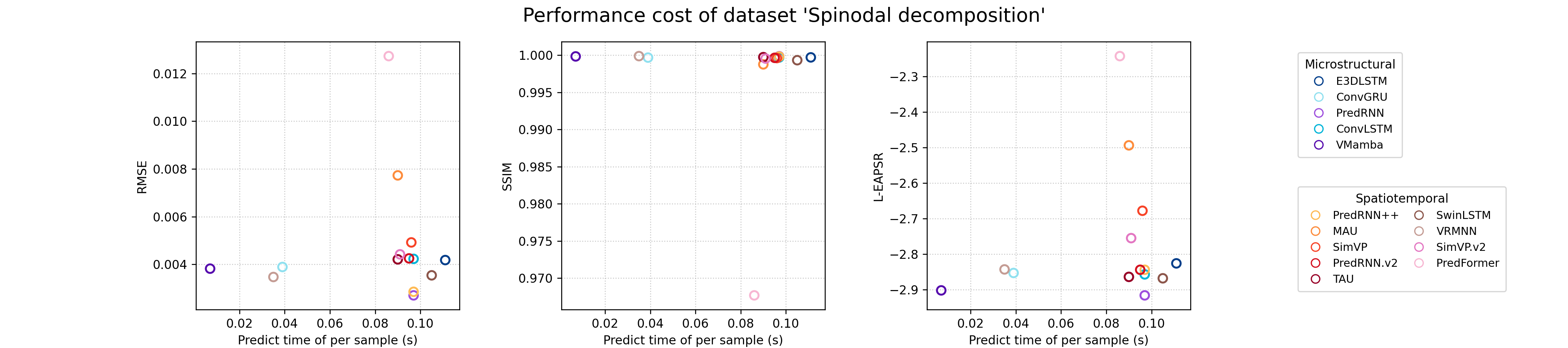}
  \caption{The performance-cost of Spinodal decomposition}
  \label{fig_perf_spin}
\end{figure*}

\begin{figure*}[ht]
  \centering
  \includegraphics[width=1\linewidth]{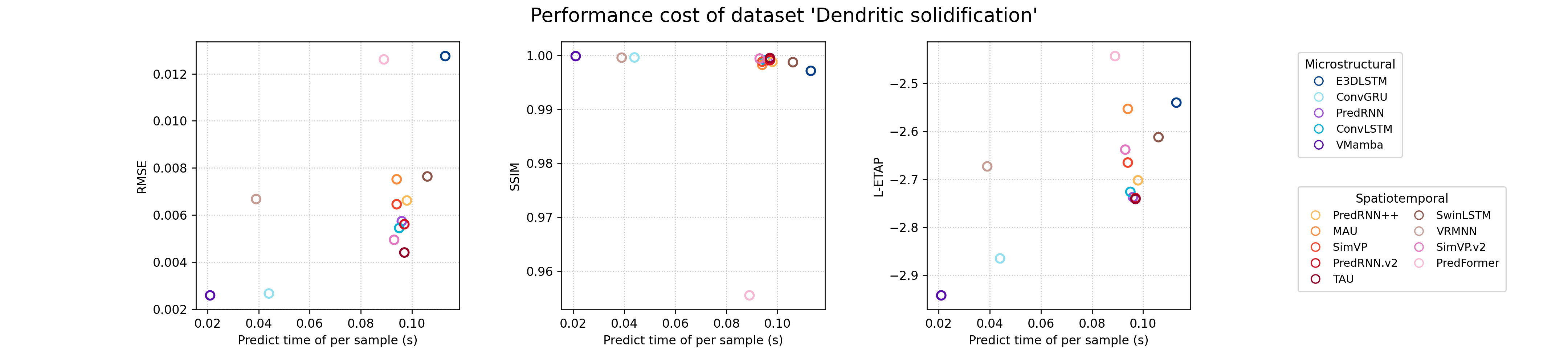}
  \caption{The performance-cost of Dendritic Solidification}
  \label{fig_perf_den}
\end{figure*}

\section{More Results}
\label{app:More Results}
In this section, we expand upon the main experimental results by presenting a more detailed view of model performance and cases on our benchmark, along with additional analysis.

\subsection{More Details of the Main Results}
\label{app:More Details of the Main Results}

This part expands upon the main results presented in the main text. We provide detailed performance tables that include supplementary numerical accuracy metrics, Mean Absolute Error (MAE) and Mean Squared Error (MSE), across all tasks and models.  In addition to these comprehensive results, we discuss some other findings and observations that complement the primary conclusions highlighted in the main body of the paper.

For each frame, Mean Absolute Error (MAE) measures the average of the absolute errors and is defined as:
\begin{equation}
\label{eq:mae}
\text{MAE} = \frac{1}{N_x N_y} \sum_{i=1}^{N_x} \sum_{j=1}^{N_y} \left| p_g(i,j) - p_p(i,j) \right|,
\end{equation}
where $p_g(i,j)$ and $p_p(i,j)$ are the pixel values of the ground truth $g$ and prediction $p$, respectively. $N_x, N_y$ are the width and height of the image in pixels.

For each frame, Mean Squared Error (MSE) measures the average of the squares of the errors and is defined as:
\begin{equation}
\label{eq:mse}
\text{MSE} = \frac{1}{N_x N_y} \sum_{i=1}^{N_x} \sum_{j=1}^{N_y} \left(p_g(i,j) - p_p(i,j)\right)^2,
\end{equation}
where the symbols have the same meaning as above. The final metric scores are computed by averaging the frame-wise results over all predicted frames and test samples.


\begin{figure*}[ht]
  \centering
  \includegraphics[width=0.7\linewidth]{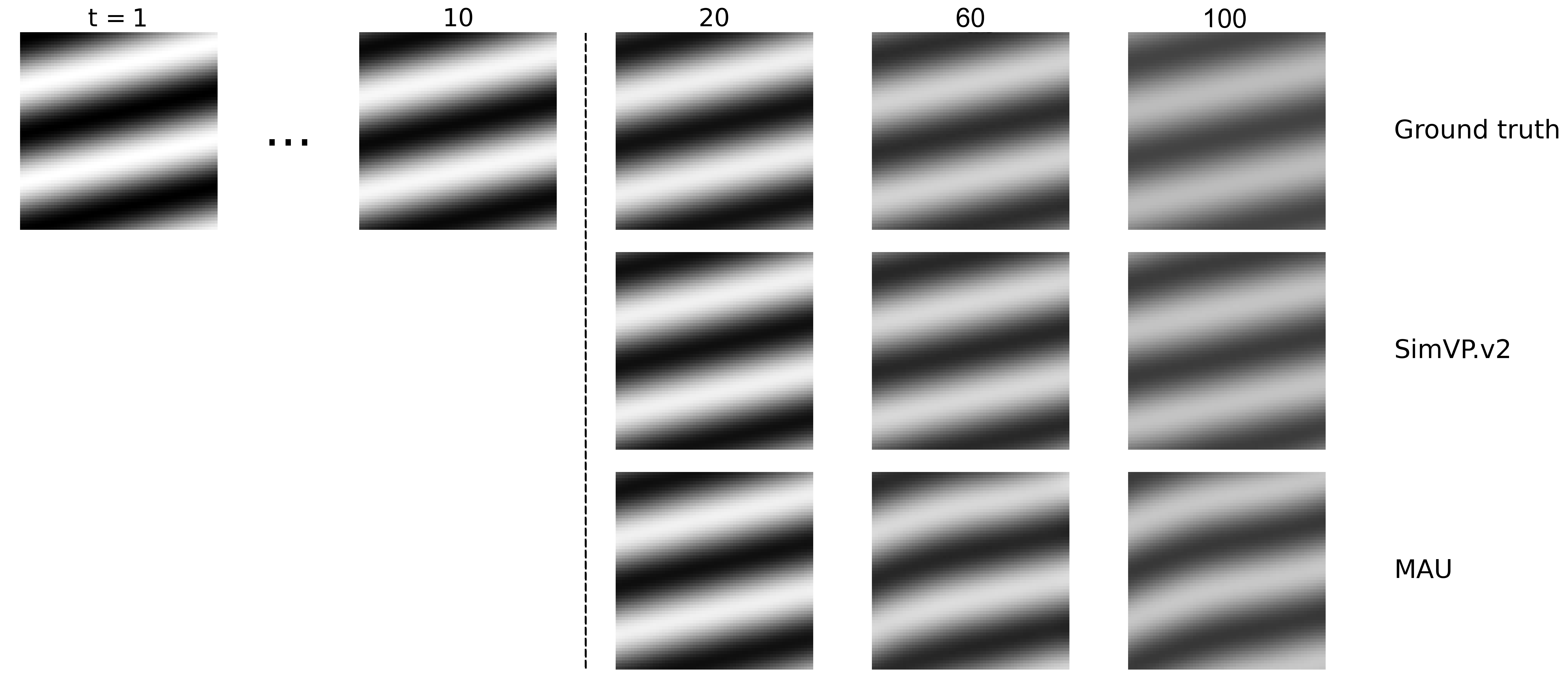}
  \caption{A case study on the plane-wave propagation. The left side shows the input sequence. The right side compares three rows: the ground truth evolution, the predictions from the high-performing model, and the predictions from the model with characteristic errors.}
  \label{fig_plane}
\end{figure*}

\textbf{MAE and MSE Metrics Further Confirm the Short- vs. Long-Term Gap.}
The inclusion of MAE and MSE in Table~\ref{tab:More Details of the Main Results} (with variance, Table~\ref{tab:Cross-Sample Variance of the Main Results}) reinforces the key finding that short-term performance is a poor indicator of long-term stability. As seen in the table, both MAE and MSE values increase by one to two orders of magnitude for nearly all models when shifting from the 10-frame to the 90-frame prediction horizon. Furthermore, we observe subtle divergences between MAE and MSE rankings. For instance, in the 90-frame prediction for Spinodal Decomposition, \textbf{TAU} achieves the best MAE (0.00310), while \textbf{VMamba} secures the best MSE (0.00047). This is because MAE (L1 norm) is less sensitive to large, outlier pixel errors than MSE (L2 norm). 

\textbf{Physical Metrics Reveal Nuances in Task Difficulty for DL Models.}
An interesting secondary finding is that our physical fidelity metrics appear to reflect the difficulty of the task for deep learning models in a way that standard numerical metrics do not. While numerical metrics like RMSE are broadly comparable across tasks, the physical fidelity scores reveal a stark contrast. Specifically, the best L-ETAP score for the Plane-wave Propagation task (around -0.99) is significantly higher (worse) than the best scores for Spinodal Decomposition (around -2.9) or Dendritic Solidification (around -2.9). This is counterintuitive, as the governing physics of plane-wave propagation are arguably the simplest. 
We hypothesize that this is because image-based DL models struggle more with the blurry, continuous interfaces of the plane-wave task than with the sharper, more defined interfaces present in the other systems. On a positive note, this suggests that the difficulty for a DL model is not necessarily correlated with the complexity of the underlying PDE. This implies that data-driven methods have the potential to effectively model physically complex systems where traditional numerical solutions are intractable, provided the visual representations of the dynamics are learnable.

Except these results from the Table, we also use Figure~\ref{fig_perf_plane}-\ref{fig_perf_den} to show the relationship of models between performance and cost, which further confirm the best results of \textbf{VMamba}.

\subsection{More Cases}
\label{app:more cases}

Here, we conduct visual case studies for each task, comparing two representative models to illustrate common performance patterns. These patterns further illustrate the challenges associated with long-term prediction and the preservation of physical properties, while also highlighting the potential benefits of introducing novel architectures. We selected one model that demonstrates strong predictive performance and another that exhibits typical failure modes. We emphasize that these models are chosen for their representativeness, not as the absolute best or worst performers. 

\begin{figure*}[ht]
  \centering
  \includegraphics[width=0.7\linewidth]{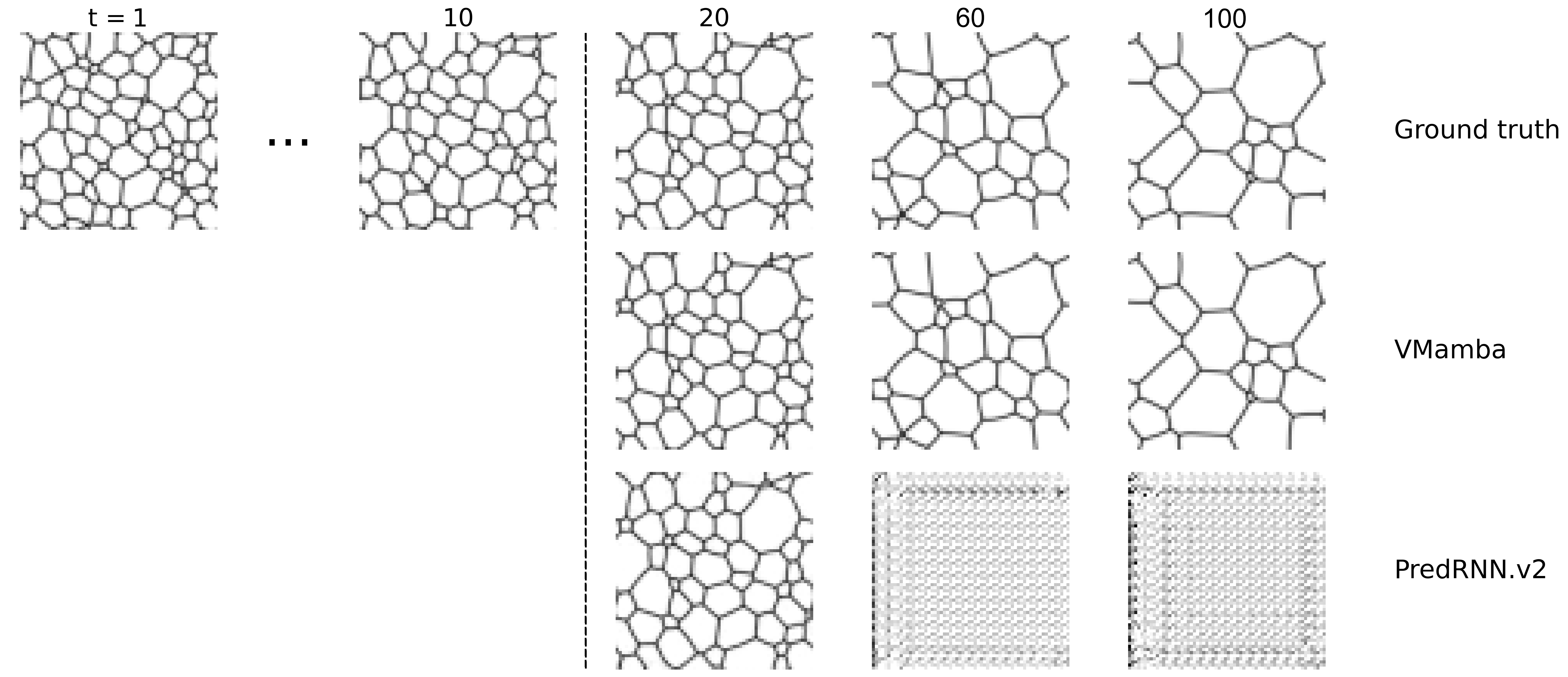}
  \caption{A case study on the grain growth. The left side shows the input sequence. The right side compares three rows: the ground truth evolution, the predictions from the high-performing model, and the predictions from the model with characteristic errors.}
  \label{fig_grain}
\end{figure*}

\begin{figure*}[ht]
  \centering
  \includegraphics[width=0.7\linewidth]{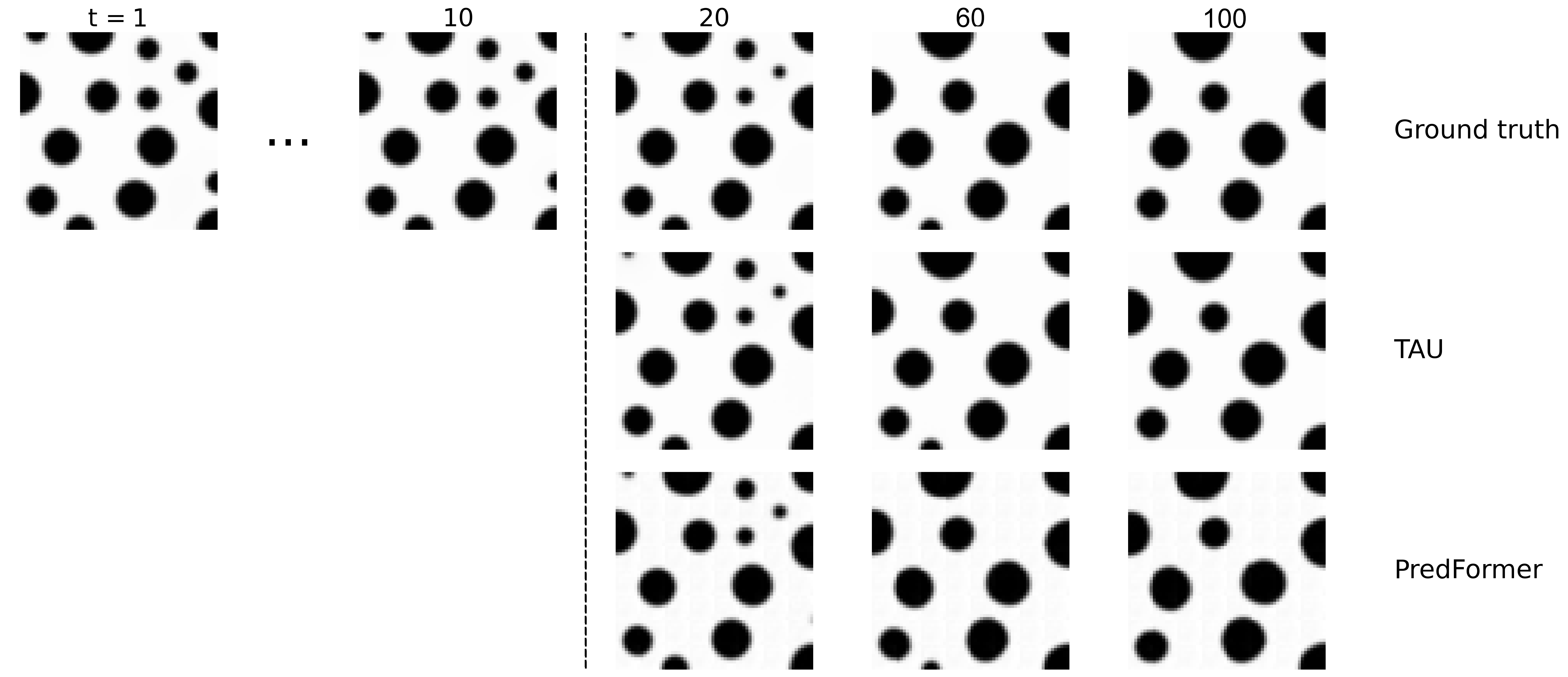}
  \caption{A case study on the spinodal decomposition. The left side shows the input sequence. The right side compares three rows: the ground truth evolution, the predictions from the high-performing model, and the predictions from the model with characteristic errors.}
  \label{fig_spin}
\end{figure*}

For the Plane-wave Propagation task (Figure \ref{fig_plane}), we compare SimVP.v2 and MAU. \textbf{This case study highlights the divergence between short-term accuracy and long-term structural integrity.}  While both models accurately capture the initial wave propagation over a short horizon, their long-term predictions differ significantly. The MAU model fails to maintain the wave, exhibiting a progressive phase distortion that causes the wave patterns to bend and lose coherence. In contrast, SimVP.v2 preserves the geometric structure of the plane wave with much higher fidelity throughout the extended forecast. This  failure of MAU to maintain structural integrity explains its significant degradation in both SSIM and L-ETAP scores. This demonstrates that even for simple dynamics, the long-term stability is still hard for some architectures.

The Grain Growth case study (Figure \ref{fig_grain}), comparing VMamba and PredRNN.v2, also \textbf{illustrates the importance of capturing the evolution of fine-grained topological features for the long-term prediction.} Both models successfully predict the general coarsening trend in the short term. However, in long-term predictions, a clear distinction emerges. The PredRNN.v2 model exhibits an accelerated and often physically inaccurate annihilation of smaller grains, failing to preserve the correct distribution and morphology. This leads to a catastrophic accumulation of error, also reflected in its exceptionally poor long-term SSIM and L-EASPR scores. Conversely, VMamba demonstrates a superior ability to accurately model the complex topological changes inherent in grain growth, maintaining high fidelity in the grain structure over the entire prediction horizon. This case highlights the strength of modern architectures like VMamba.

\begin{figure*}[ht]
  \centering
  \includegraphics[width=0.7\linewidth]{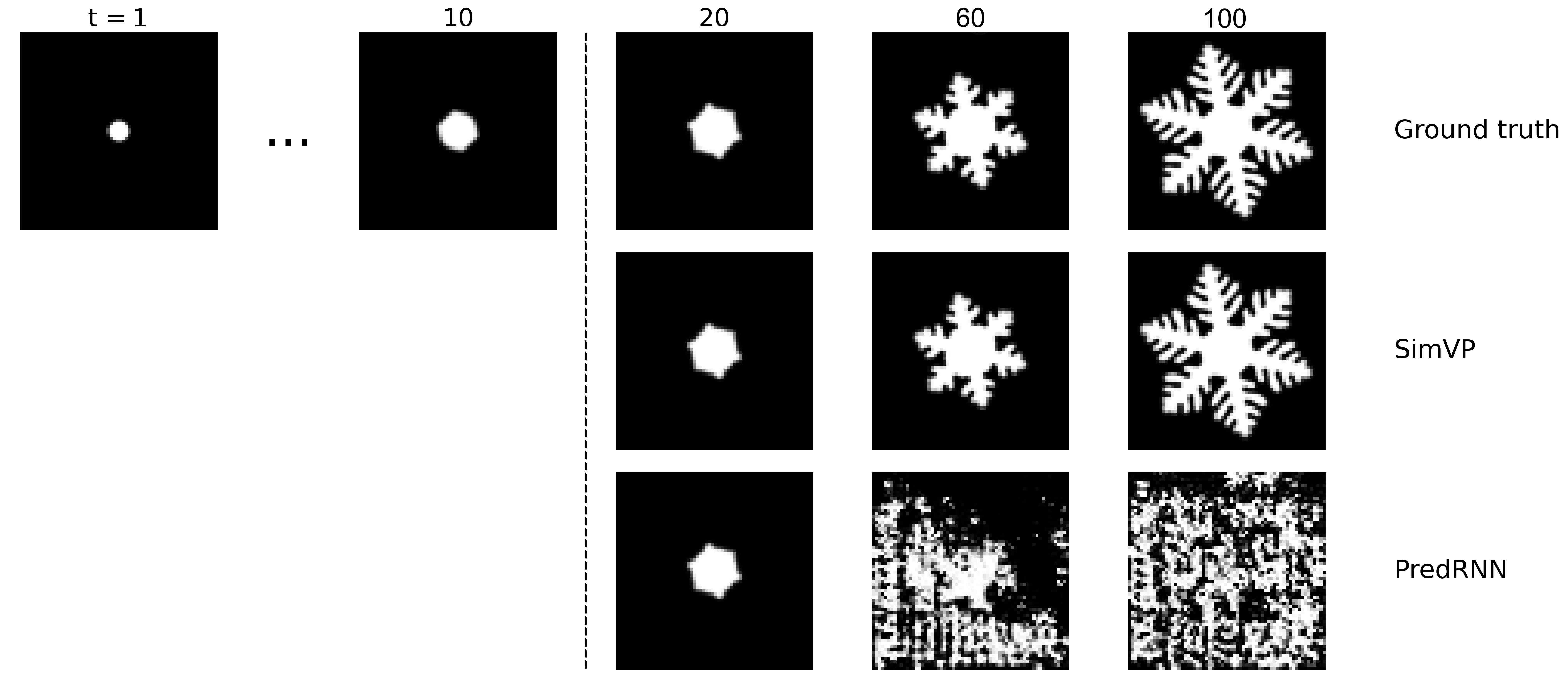}
  \caption{A case study on the dendritic solidification. The left side shows the input sequence. The right side compares three rows: the ground truth evolution, the predictions from the high-performing model, and the predictions from the model with characteristic errors.}
  \label{fig_den}
\end{figure*}

We compare two general-purpose models, TAU and PredFormer, on the Spinodal Decomposition task (Figure \ref{fig_spin}). \textbf{This case provides a nuanced look at geometric fidelity.} Both models correctly capture the high-level physics of phase coarsening, such as the dissolution of the three smaller droplets in the upper right. However, a closer inspection of their long-term predictions reveals subtle but important differences. The PredFormer model struggles to maintain precise geometric shapes; by frame 90, its predicted droplets show minor distortions and deviations from the circularity seen in the ground truth. While seemingly small, this geometric imprecision is what leads to its slightly lower SSIM and physical fidelity scores compared to TAU, which better preserves the sharp, circular interfaces. This case demonstrates that even among capable architectures, subtle design differences can impact the model's ability to adhere to precise physical characteristics.

For Dendritic Solidification (Figure \ref{fig_den}), we compare SimVP and PredRNN, \textbf{underscores the ability of modern general-purpose architectures.} Even though PredRNN was previously adapted for such tasks, it is outperformed by the newer SimVP model. While PredRNN produces reasonable short-term forecasts, it fails in the long term as the dendritic structure grows more complex. Its predictions suffer from a loss of high-frequency detail, failing to handle the huge noise. In contrast, SimVP better maintains the overall dendritic morphology and captures the fine-tipped branches with greater accuracy. However, even SimVP is not perfect, as it exhibits some blurring at the interface. This case illustrates that modern architectures are often better used to handle physical simulations, but also that even models with better performance can have characteristic failure mode.

\subsection{Main Results of Dumbbell Dataset}
\label{app:dumbbell}
To evaluate performance on our new task, we selected representative models from both the microstructural category (PredRNN, VMamba) and the spatiotemporal category (SimVP.v2). The detailed configuration of the dataset can be found in Appendix~\ref{Implementation Details of Dumbbell Dataset}. Due to the unique nature of the dumbbell evolution, which primarily focuses on the speed of motion rather than phase fraction change or grain coarsening, the existing physical fidelity metrics are not directly applicable. Therefore, our analysis for this task focuses on numerical accuracy metrics.

\textbf{The Short- vs. Long-Term Performance Gap Persists.}
Consistent with our primary results, the Dumbbell dataset reaffirms the significant gap between short- and long-term prediction accuracy. For instance, as shown in Table \ref{tab:dumbbell_results}, the RMSE for the PredRNN model degrades from 0.000415 in the short-term (10-10) prediction to 0.030105 in the long-term (10-90) prediction. This trend holds for all tested models, providing external confirmation that short-term accuracy is an unreliable indicator of a model's long-term stability.

\textbf{Nuanced Performance of VMamba Reveals Task-Specific Strengths.}
Interestingly, this dataset highlights a scenario where the otherwise dominant VMamba model shows a less pronounced advantage. While it still performs well, it is consistently outperformed by SimVP.v2 in terms of numerical accuracy across all prediction horizons. We hypothesize that this is because the dumbbell evolution involves visually simpler geometric structures compared to tasks like grain growth. The primary challenge here is not modeling complex topological changes, but rather accurately capturing the velocity of the interface evolution. In such a scenario, the sophisticated architectural priors of VMamba may not offer a distinct advantage over a highly efficient, convolution-based model like SimVP.v2, which excels at learning more direct pixel-level transformations. This suggests that even state-of-the-art architectures have specific conditions under which they are most effective.

\begin{table}[htbp]
\centering

  \caption{Main Results of Dumbbell Dataset}
    \resizebox{1\linewidth}{!}{\begin{tabular}{cccccc}
    \toprule
          &       & \multicolumn{4}{c}{Dumbbell}  \\
    \midrule
    Model domain & Model & MAE & MSE  & RMSE & SSIM  \\
    \midrule
    \multicolumn{6}{c}{Short term prediction (10-10)} \\
    \midrule
    \multicolumn{1}{c}{\multirow{2}[1]{*}{Microstructural}} & 
PredRNN & 0.000087 & 0.000000 & 0.000415 & 0.999993 \\
 & VMamba & 0.000346 & 0.000000 & 0.000548 & 0.999973 \\
    \midrule
    \multicolumn{1}{c}{\multirow{1}[1]{*}{Spatiotemporal}} & SimVP.v2 & 0.000063 & 0.000000 & 0.000229 & 0.999997 \\
    \midrule
    \multicolumn{6}{c}{Long term prediction (10-50)} \\
    \midrule
    \multicolumn{1}{c}{\multirow{2}[1]{*}{Microstructural}} & 
PredRNN & 0.000920 & 0.000116 & 0.010777 & 0.997161 \\
 & VMamba & 0.000469 & 0.000004 & 0.001897 & 0.999776 \\

    \midrule
    \multicolumn{1}{c}{\multirow{1}[1]{*}{Spatiotemporal}} & SimVP.v2 & 0.000100 & 0.000000 & 0.000679 & 0.999977 \\

    \midrule
    \multicolumn{6}{c}{Long term prediction (10-90)} \\
    \midrule
    \multicolumn{1}{c}{\multirow{2}[1]{*}{Microstructural}} & 
    PredRNN & 0.002772 & 0.000906 & 0.030105 & 0.986203\\
    &  VMamba & 0.000725 & 0.000042 & 0.006488 & 0.998383 \\
    \midrule
    \multicolumn{1}{c}{\multirow{1}[1]{*}{Spatiotemporal}} & SimVP.v2 & 0.000198 & 0.000006 & 0.002472 & 0.999717 \\
    \bottomrule
    \end{tabular}}
  \label{tab:dumbbell_results}%
\end{table}%

\subsection{Results of Out of Distribution Experiment}
\label{app:ood}

To assess the generalization capabilities of the models beyond the training distribution, we conducted an out-of-distribution (OOD) experiment. This involved testing the pre-trained models on a version of the Dumbbell dataset with physical parameters shifted outside the range seen during training, as detailed in Appendix~\ref{Implementation Details of Dumbbell Dataset}.

\textbf{The results highlight that performance degradation under OOD conditions is a persistent challenge.} As shown in Table \ref{tab:OOD_dumbbell_results}, all models experienced an increase in numerical error when faced with this distributional shift. For instance, while SimVP.v2 remains the top-performing model, its long-term (10-50) RMSE nearly doubles from 0.000679 on the in-distribution test set (Table \ref{tab:dumbbell_results}) to 0.001379 on the OOD test set (Table \ref{tab:OOD_dumbbell_results}). This performance drop, even on a task with relatively simple dynamics, underscores that out-of-distribution generalization is a fundamental challenge for current data-driven models. Therefore, developing models with improved physical grounding and resilience to distributional shifts remains a critical and open direction for future research in data-driven materials science.

\begin{table}[htbp]

\centering

  \caption{OOD Results of Dumbbell Dataset}
    \resizebox{1\linewidth}{!}{\begin{tabular}{cccccc}
    \toprule
          &       & \multicolumn{4}{c}{Dumbbell}  \\
    \midrule
    Model domain & Model & MAE & MSE  & RMSE & SSIM  \\
    \midrule
    \multicolumn{6}{c}{Short term prediction (10-10)} \\
    \midrule
    \multicolumn{1}{c}{\multirow{2}[1]{*}{Microstructural}} & 
PredRNN & 0.000118 & 0.000002 & 0.001350 & 0.999987 \\
 & VMamba & 0.000448 & 0.000001 & 0.001140 & 0.999943 \\
    \midrule
    \multicolumn{1}{c}{\multirow{1}[1]{*}{Spatiotemporal}} & SimVP.v2 & 0.000090 & 0.000000 & 0.000548 & 0.999994 \\
    \midrule
    \multicolumn{6}{c}{Long term prediction (10-50)} \\
    \midrule
    \multicolumn{1}{c}{\multirow{2}[1]{*}{Microstructural}} & 
PredRNN & 0.000779 & 0.000096 & 0.009812 & 0.997378 \\
 & VMamba & 0.000552 & 0.000006 & 0.002449 & 0.999696 \\

    \midrule
    \multicolumn{1}{c}{\multirow{1}[1]{*}{Spatiotemporal}} & SimVP.v2 & 0.000156 & 0.000002 & 0.001379 & 0.999915 \\

    \midrule
    \multicolumn{6}{c}{Long term prediction (10-90)} \\
    \midrule
    \multicolumn{1}{c}{\multirow{2}[1]{*}{Microstructural}} & 
    PredRNN & 0.002061 & 0.000572 & 0.023913 & 0.990124\\
    &  VMamba & 0.000725 & 0.000042 & 0.006488 & 0.998383 \\
    \midrule
    \multicolumn{1}{c}{\multirow{1}[1]{*}{Spatiotemporal}} & SimVP.v2 &  0.000265 & 0.000008 & 0.002782 & 0.999598 \\
    \bottomrule
    \end{tabular}}
  \label{tab:OOD_dumbbell_results}%
\end{table}%

\section{Future Directions and Limitations}
\label{app:limitation}

\paragraph{Future Directions}
Building upon the insights from our benchmark, we identify two another compelling directions for future research in this domain.
\begin{itemize}
    \item \textbf{Models with Physical Informations.} While our work demonstrates the superiority of modern architectures like state-space models, a critical next step is to more deeply integrate physical priors directly into these powerful frameworks. This goes beyond post-hoc evaluation with physical metrics. Future research could explore models inductive biases that inherently respect physical laws such as mass conservation or thermodynamic principles. This would bridge the gap between high-performance generalist architectures and the specific physical constraints, leading to more robust and trustworthy models.
    \item \textbf{Foundational Models for Microstructure Evolution.} Inspired by the success of Large Language Models (LLMs) and other foundational models, a paradigm shift from task-specific models towards a unified model for microstructure evolution is a highly promising avenue. Such a foundational model could be pre-trained on a massive and diverse corpus of microstructural data, encompassing different processes and physical phenomena. Once pre-trained, it could be fine-tuned to excel at a wide array of downstream tasks, potentially demonstrating strong zero-shot or few-shot generalization capabilities.
\end{itemize}

\paragraph{Limitations}
We acknowledge certain limitations in the current scope of {\tt MicroEvoEval}. Primarily, the generation of high-fidelity microstructural data via numerical methods is exceptionally time-consuming. Consequently, this work has focused on curating, standardizing, and reprocessing datasets for several classic and representative MicroEvo tasks, rather than generating entirely new ones. While our benchmark provides a robust foundation, we plan to expand its coverage in the future by generating and incorporating data for a broader range of tasks and material systems, further enhancing its comprehensiveness and utility for the community.

\clearpage
\newpage
This is the \textbf{Code and Data Appendix} of our paper. In this part, we will give more details of how the benchmark is constructed and the implementation details of models for reproducibility.

\section{Implementation Details of Our Benchmark}

Here, we further introduce all the implementation details. The corresponding data and code can be found in the main paper. Our full datasets and the code of our benchmark are publicly available on Hugging Face and GitHub with \textbf{cc-by-nc-4.0 and Apache-2.0 license}, respectively. We are committed to maintaining and updating them based on community feedback and our future research.

\subsection{Implementation Details of Datatsets}
\label{app:Implementation Details of Datatsets}

In this part, we describe the raw data and the specific preprocessing steps we applied.

\paragraph{Original Datasets.} We built our {\tt MicroEvoEval} framework based mainly on four publicly available datasets from~\cite{yang2021self} as our raw datasets (can be accessed in their original paper). This is primarily because regenerating the entire dataset with numerical methods is extremely time-consuming. Furthermore, as we state in the main text, these datasets already cover representative microstructural evolution tasks we want to evaluate: plane-wave propagation, grain growth, spinodal decomposition, and dendritic solidification.

\begin{itemize}
\item \textbf{Plane-Wave Propagation.} The original dataset is split into a training set with 240 samples (each containing 200 frames), a validation set with 30 samples (200 frames each), and a test set with 100 samples (200 frames each).

\item \textbf{Grain Growth.} The original dataset is split into a training set with 30,400 samples (each containing 1 frame, extracted from longer sequences constructed by concatenating multiple 20-frame clips), a validation set with 7,600 samples (1 frame each, extracted in the same way), and a test set with 1,000 samples (each containing 200 frames) representing full grain growth sequences.

\item \textbf{Spinodal Decomposition.} The original dataset includes a training set with 640 samples (100 frames each), a validation set with 159 samples (100 frames each), and a test set with 510 samples (201 frames each).

\item \textbf{Dendritic Solidification.} This original dataset consists of a training set with 940 samples (100 frames each), a validation set with 60 samples (100 frames each), and a test set with 100 samples (100 frames each).

\end{itemize}

All images contained in the above dataset are the height of 64 pixels
and width of 64 pixels of each frame, and all statistical information is shown in Table~\ref{tab:dataset}.

\begin{table*}[htbp]
\renewcommand{\arraystretch}{1.2}
\setlength{\tabcolsep}{0.5mm} 
  \centering
  \small
    \caption{ The Statistic of the Original Datasets. The datasets are represented as tensors with the shape $\mathbf{(N,T,H,W)}$, where N denotes the number of samples, T represents the frame size of a sample, and H and W correspond to the height and width of each frame, respectively. }

\begin{tabular}{ccccc}
\toprule

\textbf{Datasets} & \textbf{Plane-Wave propagation} & \textbf{Grain Growth} & \textbf{Spinodal Decomposition} & \textbf{Dendrite Solidification} \\ 
         \midrule
\textbf{Train} & (240, 200, 64, 64) & (30400, 1, 64, 64) & (640, 100, 64, 64) & (940, 100, 64, 64) \\ 
        \midrule
\textbf{Valid} & (30, 200, 64, 64)  & (7600, 1, 64, 64)  & (159, 100, 64, 64) & (60, 100, 64, 64)  \\ 
        \midrule
\textbf{Test}  & (100, 200, 64, 64) & (1000, 200, 64, 64) & (510, 201, 64, 64) & (100, 100, 64, 64) \\ 
        \bottomrule
        \end{tabular}
        \label{tab:dataset} 
\end{table*}

\paragraph{Our Dataset.}
To support our benchmark's investigation into both short- and long-term performance across a variety of metrics, we performed extensive processing on the original dataset to construct our own benchmark dataset based on their splitting of train, validation and test sets.
We apply the tailored preprocess as follows:

\begin{itemize}
    \item \textbf{Training Sets.} For the \textbf{plane-wave propagation}, \textbf{spinodal decomposition}, and \textbf{dendritic solidification} datasets, we employ a sliding window of length 20 (coresponding to the 10-10, forecasting \texttt{10} frames from \texttt{10} input frames) with a stride of 4 to generate training and validation samples. This provides sufficient data volume for training long-term prediction models while maintaining computational efficiency.     
    For the \textbf{grain growth} dataset, due to its unique structure, where frames are not grouped by sample and each sample contains only 20 frames, we apply a non-overlapping sliding window of length 20 to generate training and validation sequences.

    \item \textbf{Validation Sets.} The validation set is processed identically to the training set. Strictly speaking, the validation set is not a direct component of our benchmark; its primary role is to serve as a reference for epoch selection during the training of each model we evaluate.
    
    \item \textbf{Test Sets.} We apply a sliding window of lengths 20, 60, and 100, corresponding to short-term (10-10) and long-term (10-50, 10-90) prediction tasks, respectively. The stride is uniformly set to 4. 
\end{itemize}



\paragraph{Data Format and Structure.}
In our benchmark, for each dataset, the training, validation, and test sets are stored in five (one for training, one for validation, one for 10-10 short-term test, one for 10-50 long-term test, and one for 10-90 long-term test) separate \textit{NumPy} archive files with the \texttt{.npy} extension. Each file behaves like a dictionary and is named according to the name of dataset and frames (like 10-10). In total, the full benchmark includes 20 dataset files. Due to the extensive size of our processed dataset (79G), we are only able to upload a sampled test dataset for each task in the supplementary materials. We promise that our full datasets of our benchmark will be made publicly available on Hugging Face after official publication.



\begin{figure*}[htbp]
  \centering
  \includegraphics[width=0.7\linewidth]{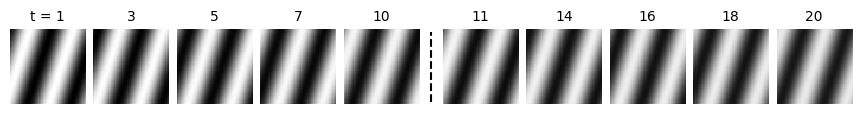}
  \caption{A sample of training set of Plane-wave propagation dataset with 10 input frames and 10 output frames.}
  \label{fig_plane_sample_train}
\end{figure*}

\begin{figure*}[htbp]
  \centering
  \includegraphics[width=0.7\linewidth]{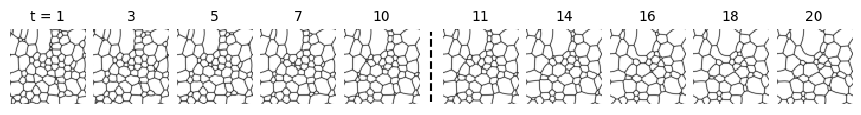}
  \caption{A sample of training set of  Grain growth dataset with 10 input frames and 10 output frames.}
  \label{fig_grain_sample_train}
\end{figure*}

\begin{figure*}[htbp]
  \centering
  \includegraphics[width=0.7\linewidth]{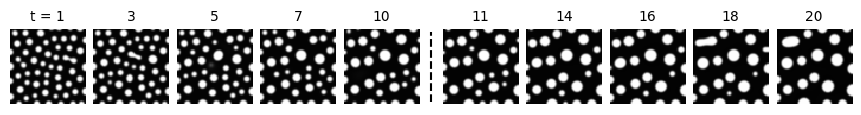}
  \caption{A sample of training set of  Spinodal decomposition dataset with 10 input frames and 10 output frames.}
  \label{fig_spin_sample_train}
\end{figure*}

\begin{figure*}[htbp]
  \centering
  \includegraphics[width=0.7\linewidth]{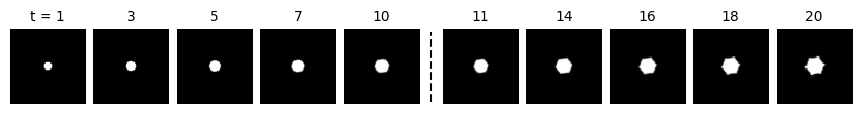}
  \caption{A sample of training set of   Dendritic solidification dataset with 10 input frames and 10 output frames.}
  \label{fig_den_sample_train}
\end{figure*}

\begin{figure*}[htbp]
  \centering
  \includegraphics[width=0.7\linewidth]{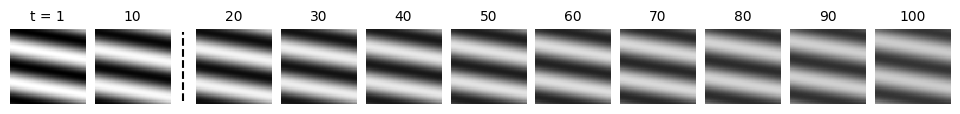}
  \caption{A sample of test set of Plane-wave propagation dataset with 10 input frames and 90 output frames.}
  \label{fig_plane_sample}
\end{figure*}

\begin{figure*}[htbp]
  \centering
  \includegraphics[width=0.7\linewidth]{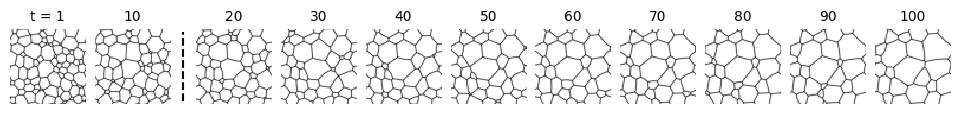}
  \caption{A sample of test set of Grain growth dataset with 10 input frames and 90 output frames.}
  \label{fig_test_sample}
\end{figure*}

\begin{figure*}[htbp]
  \centering
  \includegraphics[width=0.7\linewidth]{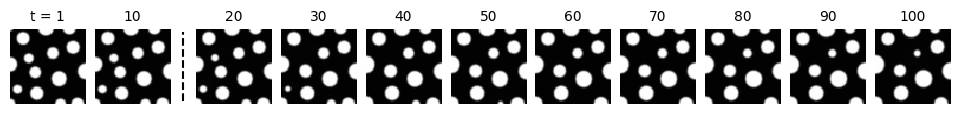}
  \caption{A sample of test set of  Spinodal decomposition dataset with 10 input frames and 90 output frames.}
  \label{fig_spin_sample}
\end{figure*}

\begin{figure*}[htbp]
  \centering
  \includegraphics[width=0.7\linewidth]{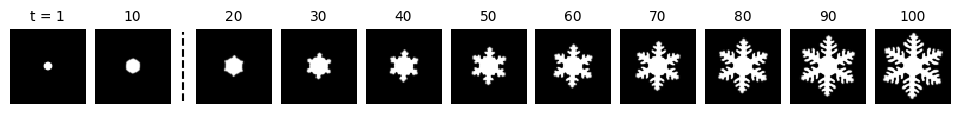}
  \caption{A sample of test set of   Dendritic solidification dataset with 10 input frames and 90 output frames.}
  \label{fig_den_sample}
\end{figure*}

\paragraph{Data Instances.} Here, we show some data instances from the training (10-10) and test (10-90) sets of the four datasets in our benchmark (Figure \ref{fig_plane_sample_train} - \ref{fig_den_sample}). Some intermediate frames are omitted for brevity.



\begin{table*}[htbp]
\renewcommand{\arraystretch}{1.2}
\setlength{\tabcolsep}{0.5mm} 
  \centering
  \small

    \caption{ The Statistics of the Training, Validation and Test Dataset in Our {\tt MicroEvoEval} Benchmark.}

        \begin{tabular}{ccccc}
        \toprule

\textbf{Datasets} & \textbf{Plane-Wave Propagation} & \textbf{Grain Growth} & \textbf{Spinodal Decomposition} & \textbf{Dendrite Solidification} \\ 
\midrule
\textbf{Train(10-10)}       & 11040                   & 1520                   & 13440                  & 19740                   \\
\midrule
\textbf{Valid(10-10)}       & 1380                    & 380                     & 3339                   & 1260                    \\
\midrule
\textbf{Test(10-10)} & 4600                     & 46000                   & 23460                  & 2100                     \\
\midrule
\textbf{Test(10-50)} & 3600                     & 36000                   & 18360                  & 1100                     \\
\midrule
\textbf{Test(10-90)} & 2600                     & 26000                   & 13260                  & 100    \\

        \bottomrule
        \end{tabular}
    \label{tab:ourdataset}
\end{table*}

\paragraph{Data Statistics.} In our benchmark, the overall statistics of the datasets are shown in Table \ref{tab:ourdataset}. To describe it in the text, the number of training samples is as follows: 11,040 for the \textbf{Plane-Wave Propagation} dataset, 13,400 for the \textbf{Spinodal Decomposition} dataset, 1,520 for the \textbf{Grain Growth} dataset, and 19,740 for the \textbf{Dendritic Solidification} dataset.
The number of test samples varies across tasks and datasets due to the short- and long-term prediction. 
\begin{itemize}
    \item \textbf{Short-term Prediction (10-10):} The number of test samples is 4,600 for Plane-Wave Propagation, 23,460 for Spinodal Decomposition, 4,600 for Grain Growth, and 2,100 for Dendritic Solidification.
    
    \item \textbf{Long-term Prediction (10-50):} The number of test samples is 3,600 for Plane-Wave Propagation, 18,360 for Spinodal Decomposition, 36,000 for Grain Growth, and 1,100 for Dendritic Solidification.
    
    \item \textbf{Long-term Prediction (10-90):} The number of test samples is 2,600 for Plane-Wave Propagation, 13,260 for Spinodal Decomposition, 26,000 for Grain Growth, and 100 for Dendritic Solidification.
\end{itemize}
Due to the extensive size of our processed dataset (79G), we are only able to upload a sampled test dataset for each task in the supplementary materials. We promise that our full datasets of our benchmark will be made publicly available on Hugging Face after official publication.

\paragraph{Dataset Curators.}
The datasets are created by all authors (will be public available after the double-blind revision). They bear all responsibility for these datasets. 

\paragraph{Licensing Information.} The datasets will be under the cc-by-nc-4.0 license.

\begin{table*}[htbp]
\footnotesize
\renewcommand{\arraystretch}{1.5} 
\centering
\caption{The Models Evaluated in Our {\tt MicroEvoEval} Benchmark.}
\begin{tabular}{ccc}
\toprule
{\bf Model Domain}      & {\bf Model}              & {\bf Conference~/~Journal}                   \\
\midrule
 \multicolumn{1}{c}{\multirow{5}[2]{*}{Microstructural}}   & E3D-LSTM~\cite{yang2021self}  & Patterns 2021                            \\ 
   & ConvGRU~\cite{lanzoni2022morphological}& Phys. Rev. Mater. 2022         \\ 
   & PredRNN~\cite{farizhandi2023spatiotemporal}  & Comput. Mater. Sci.2023                         \\
  & ConvLSTM~\cite{mao2024spatiotemporal}& Mater. Today Commun. 2024                       \\ 

    & VMamba~\cite{jing2025research} & Comput. Mater. Sci. 2025   \\ 
\midrule

\multicolumn{1}{c}{\multirow{9}[2]{*}{Spatiotemporal}} & PredRNN++~\cite{Wang2018predrnn++} & ICML2018                           \\ 
        &MAU~\cite{chang2021mau} & NeurIPS 2021                           \\ 
    & SimVP~\cite{gao2022simvp}     &CVPR 2022                    \\ 
& PredRNN.v2~\cite{wang2022predrnn} & TPAMI 2022                            \\

      & TAU~\cite{tan2023temporal}  & CVPR 2023                              \\ 
  & SwinLSTM~\cite{tang2023swinlstm}  & ICCV 2023                             \\ 
     & VMRNN~\cite{tang2024vmrnn} & CVPR2024                               \\ 
  &SimVP.v2~\cite{tan2025simvpv2} & TMM 2025       \\ 
 & PredFormer~\cite{tang2024predformer} & Arxiv 2025                        \\ 
\bottomrule
\end{tabular}
\label{tab:model_info}
\end{table*}

\subsection{Implementation Details of Metrics}
\label{app:Implementation Details of metrics}
In this part, we describe how the metrics for our benchmark were implemented.

\paragraph{The Metrics.}
In our benchmark, we totally have four numerical accuracy metrics, two physical fidelity metrics and the computational efficiency in the main text and Appendix.

\begin{itemize}
    \item \textbf{Numerical Accuracy Metrics}
    \begin{itemize}
    
        \item \textbf{RMSE.} Root Mean Squared Error, defined for each frame as in Equation~\eqref{eq:rmse} in the main text.
        
        \item \textbf{SSIM.} Structural Similarity Index Measure, which quantifies perceptual similarity between two images, as defined in Equation~\eqref{eq:ssim} in the main text.
        
        \item \textbf{MSE.} Mean Squared Error, computed per frame as shown in Equation~\eqref{eq:mse} in the Technical Appendix.
        
        \item \textbf{MAE.} Mean Absolute Error, also defined per frame, as in Equation~\eqref{eq:mae} in the Technical Appendix.
    \end{itemize}

    \item \textbf{Physical Fidelity Metrics}
    \begin{itemize}
        \item \textbf{L-ETAP.} Log-Error of Total Area Proportion, a physical metric designed for the \textbf{Plane-Wave Propagation}, \textbf{Spinodal Decomposition}, and \textbf{Dendritic Solidification} tasks, defined in Equation~\eqref{eq:l-etap} in the main text.
        
        \item \textbf{L-EAPSR.} Log-Error of Average Proportion of Single Region, a physical metric designed for the \textbf{Grain Growth} task, as defined in Equation~\eqref{eq:l-eapsr} in the main text.
    \end{itemize}

    \item \textbf{Computational Efficiency}
    \begin{itemize}
        \item \textbf{Inference Time.} This is quantified by the average inference time required to predict a complete short-term sequence (forecasting \texttt{10} frames from \texttt{10} input frames), as defined in the main text.
    \end{itemize}
\end{itemize}

\paragraph{More Details on Physical Fidelity Metrics.}
Since our physical fidelity metric is designed to evaluate the properties of the physical field across time, we compute the spatial $\ell_2$ norm for each frame of every sample. We then report the root mean square (RMS) of all these collected norms over the entire test set.
Formally, for a given sample $i$, let $x_{i,t}$ denote the quantity of interest at time $t$ (e.g., phase fraction or average grain size). The temporal $\ell_2$ norm of $x_{i,t}$ is defined as $\|x_{i,t} \|_{2,i,T} = \sqrt{\sum_{t=0}^T x_{i,t}^2}$.

For tasks such as \textbf{plane-wave propagation}, \textbf{spinodal decomposition}, and \textbf{dendritic solidification}, the primary physical quantity of interest is the global phase fraction, which corresponds to the proportion of the image covered by a monochromatic region. Based on this, we define the \textbf{Log-Error of Total Area Proportion (L-ETAP)} as:
\begin{align*}
\text{L-ETAP}
&= \log_{10}\left( \sqrt{\frac{1}{N} \sum_{i=1}^N \left\| \frac{A(\Omega^g_{i,t})}{S} - \frac{A(\Omega^p_{i,t})}{S} \right\|_{2,i,T}^2 } \right) \\
&= \log_{10}\left( \sqrt{\frac{1}{N} \sum_{i=1}^N \sum_{t=0}^T \left( \frac{A(\Omega^g_{i,t})}{S} - \frac{A(\Omega^p_{i,t})}{S} \right)^2 } \right),
\end{align*}
where $A(\Omega_{i,t})$ denotes the area of the monochrome region in frame $\Omega_{i,t}$, $S$ is the total image area, and superscripts $g$ and $p$ refer to ground truth and predicted frames, respectively.

For the \textbf{grain growth} task, we focus on the average crystal grain size, which is inversely related to the number of grains. We define the \textbf{Log-Error of Average Proportion of Single Region (L-EAPSR)} as:
\begin{align*}
\text{L-EAPSR}
&= \log_{10}\left( \sqrt{\frac{1}{N} \sum_{i=1}^N \left\| \frac{1}{C(\Omega^g_{i,t})} - \frac{1}{C(\Omega^p_{i,t})} \right\|_{2,i,T}^2 } \right) \\
&= \log_{10}\left( \sqrt{\frac{1}{N} \sum_{i=1}^N \sum_{t=0}^T \left( \frac{1}{C(\Omega^g_{i,t})} - \frac{1}{C(\Omega^p_{i,t})} \right)^2 } \right),
\end{align*}
where $C(\Omega_{i,t})$ denotes the number of grains (connected regions) in frame $\Omega_{i,t}$.

\noindent\textbf{Implementation of $A(\Omega_{i,t})$ and $C(\Omega_{i,t})$.}
To compute the physical quantities $A(\Omega_{i,t})$ and $C(\Omega_{i,t})$, we adopt the following strategies:

\begin{itemize}
    \item \textbf{Area computation $A(\Omega_{i,t})$:}  
    Since the images from physical simulations are grayscale, we first binarize each frame to distinguish the phase. We set the binarization threshold to half of the maximum grayscale value in the image. Pixels with intensity above the threshold are labeled white; others are black. The area is then computed by counting the number of black pixels.
    
    \item \textbf{Connected region count $C(\Omega_{i,t})$:}  
    To count the number of grains (connected components), we use OpenCV's \texttt{findContours} function. Prior to this, each image is upsampled to $256 \times 256$ pixels to improve contour resolution. We also use binarization, and morphological operations to eliminate noise, ensuring more accurate contour detection and region counting.
\end{itemize}

\paragraph{Licensing Information.} The metrics will be part of our code framework under the Apache-2.0 license. 

\subsection{Implementation Details of Dumbbell Dataset}
\label{Implementation Details of Dumbbell Dataset}
In this part, we describe the raw data and the metrics we used in this new Dumbbell dataset. Unless otherwise specified, other details will remain consistent with the other four datasets.

\paragraph{The raw data.} The statistic of the train dataset, validation dataset and test dataset are (200, 100, 64, 64), (30, 100, 64, 64) and (50, 100, 64, 64) respectively. We split the train dataset and the validation dataset into $1800$ samples (each containing 20 frames) and 270 samples (each containing 20 frames) respectively. The test dataset is splitting into 450 samples (each containing 20 frames) for short-term prediction (10-10), 200 samples (each containing 60 frames) for long-term prediction (10-50) and 50 samples (each containing 100 frames) for long-term prediction (10-90).

\paragraph{OOD dataset construction.}
To evaluate the models’ out-of-distribution generalization ability, we additionally construct an OOD version of the Dumbbell dataset with the same size of standard version by shifting the  physical parameters \textit{radius}. In the meta-dataset used for training and validation, the dumbbell radius is sampled from the interval 
$[0.31,0.39]$, in contrast, the OOD dataset is generated using a larger radius range of $[0.39,0.43]$, which lies entirely outside the training distribution.

 \paragraph{The metrics.} In this dataset, we also have the four numerical accuracy metrics that we describe in Appendix \ref{app:Implementation Details of metrics}. Due to the unique physical properties of this new dataset, we also plan to design new, corresponding physical metrics in our future work.


\section{Implementation Details of Models}
\label{app:implementation_details_of_model}
In this part, we describe how the models in our benchmark were implemented.

\paragraph{The Models in Our Benchmark.}
To ensure comprehensive coverage, we evaluate a total of 14 models, spanning both domain-specific microstructure prediction models and general-purpose spatiotemporal architectures, as summarized in Table~\ref{tab:model_info}.

    

\paragraph{Code Framework.}
We gratefully acknowledge the \textbf{OpenSTL}, which were instrumental in reproducing the models and developing our framework. It is licensed under the \textbf{Apache-2.0 license}. We adapt and extend this framework to support the training and evaluation of models for our image-based microstructure evolution, as well as to validate our proposed physical metrics.

\begin{table*}[htbp]
\renewcommand{\arraystretch}{1.2}
\setlength{\tabcolsep}{0.5mm} 
  \centering
  \small
  \caption{The Final Chosen Epoch of Each Model Evaluated in Our {\tt MicroEvoEval} Benchmark.}
    \begin{tabular}{cccccc}
   \toprule
    \textbf{Model domain} & \textbf{Model} &\textbf{Plane-Wave Propagation} & \textbf{Grain Growth}& \textbf{Spinodal Decomposition}& \textbf{Dendritic Solidification}  \\
    \midrule
    \multicolumn{1}{c}{\multirow{5}[2]{*}{Microstructural}} & E3DLSTM & 200 & 300 & 250 & 300 \\
          & ConvGRU & 200 & 300 & 250 & 400 \\
          & PredRNN & 200 & 300 & 250 & 400 \\
          & ConvLSTM & 200 & 300 & 250 & 300 \\
          & VMamba & 200 & 250 & 200 & 300 \\
    \midrule
    \multicolumn{1}{c}{\multirow{9}[2]{*}{Spatiotemporal}} & PredRNN++ & 200 & 250 & 200 & 250 \\
          & MAU    & 200 & 300 & 250 & 400 \\
          & SimVP  & 200 & 250 & 250 & 250 \\
          & PredRNN.v2  & 200 & 300 & 250 & 400 \\
          & TAU  & 200 & 300     & 250 & 400 \\
          & SwinLSTM   & 200 & 250 & 250 & 300 \\\
          & VMRNN & 200 & 300 & 200 & 400 \\    
          & SimVP.v2  & 200 & 250 & 200 & 300 \\
          & PredFormer & 50  & 200  & 50 & 50 \\
    \bottomrule
    \end{tabular}%
  \label{tab:epoch}%
\end{table*}%

\paragraph{Hyperparameters And Model Details.}
All settings follow the default configurations of each model unless otherwise specified in the following paragraphs. The primary modifications we make are as follows:
\begin{itemize}
    \item \textbf{Epochs.}  
    The max number of training epochs is determined based on the relative complexity of the physical problems in each dataset. We assign more training epochs to more complex tasks. Specifically, the ordering from lowest to highest number of epochs is: \textbf{Plane-Wave Propagation} $<$ \textbf{Grain Growth} $\approx$ \textbf{Spinodal Decomposition} $<$ \textbf{Dendritic Solidification}. Under this setting, all models were trained for a maximum of 200, 300, 300, and 400 epochs of each task with the best performing checkpoint selected based on performance in the held-out validation set. The final chosen checkpoints with epoch settings are listed in Table~\ref{tab:epoch}.

    \item \textbf{Learning Rates.} When selecting the learning rate for each model, we prioritized the default values recommended by the original paper or the OpenSTL framework. When a model's performance on the validation set was particularly poor, we manually adjust its learning rate. The specific learning rate for E3DLSTM is 0.0001, for ConvGRU is 0.0001, for PredRNN is 0.0005, for ConvLSTM is 0.0005, for VMamba is 0.001, for PredRNN++ is 0.0001, for MAU is 0.001, for SimVP is 0.001, for PredRNN.v2 is 0.0005, for TAU is 0.001, for SwimLSTM is 0.0001, for VMRNN is 0.0001, for SimVP.v2 is 0.001, and for PredFormer is 0.001.
    
    \item \textbf{Training and Prediction.}  
    Following the standard setting in the image-based microstructure evolution \cite{yang2021self}, all models are trained under the short-term prediction setting (10-10); for long-term prediction tasks (10-50 and 10-90), we keep use the autoregressive strategy, where the trained short-term model is iteratively applied to generate future frames.
\end{itemize}


\paragraph{Computation Enviroment.}
All models are trained on a Linux server equipped with 1024GB memory, an Intel(R) Xeon(R) Gold 6430 CPU and  4 $\times$ NVIDIA A6000 GPUs. 

\paragraph{Licensing Information.} The implementation of the models will be part of our code framework under the Apache-2.0 license.


\end{document}